\documentclass[conference]{IEEEtran}
\IEEEoverridecommandlockouts
\usepackage{cite}
\usepackage{amsmath,amssymb,amsfonts}
\usepackage{algorithmic}
\usepackage{graphicx}
\usepackage{textcomp}
\usepackage{xcolor}
\usepackage{subfig}
\def\BibTeX{{\rm B\kern-.05em{\sc i\kern-.025em b}\kern-.08em
    T\kern-.1667em\lower.7ex\hbox{E}\kern-.125emX}}
\usepackage{soul}

\DeclareSubrefFormat{myparens}{#1(#2)}
\begin{document}

\title{Scavenger: Better Space-Time Trade-Offs for Key-Value Separated LSM-trees\\
\thanks{* Corresponding authors}
}

\author{
    \IEEEauthorblockN{Jianshun Zhang{$^\dagger$}, Fang Wang{$^{\dagger \ddagger *}$}, Sheng Qiu{$^{\diamond}$}, Yi Wang{$^\diamond$}}
    \IEEEauthorblockN{Jiaxin Ou{$^{\diamond *}$}, Junxun Huang{$^\dagger$}, Baoquan Li{$^\dagger$}, Peng Fang{$^{\dagger *}$}, Dan Feng{$^\dagger$}}
    \IEEEauthorblockA{{$^\dagger$}Wuhan National Laboratory for Optoelectronics, Key Laboratory of Information Storage System, \\ Engineering Research Center of data storage systems and Technology, Ministry of Education of China, \\ School of Computer Science and Technology, Huazhong University of Science and Technology, China \\ {$^\ddagger$}Shenzhen Huazhong University of Science and Technology Research Institute {$^\diamond$}ByteDance Inc.}
    \IEEEauthorblockA{{\{shunzi, wangfang, junix, bqli, fangpeng, dfeng\}@hust.edu.cn,  \{sheng.qiu, wangyi.ywq, oujiaxin\}@bytedance.com}}
}
\maketitle

\begin{abstract}
Key-Value Stores (KVS) implemented with log-structured merge-tree (LSM-tree) have gained widespread acceptance in storage systems. Nonetheless, a significant challenge arises in the form of high write amplification due to the compaction process. While KV-separated LSM-trees successfully tackle this issue, they also bring about substantial space amplification problems, a concern that cannot be overlooked in cost-sensitive scenarios. Garbage collection (GC) holds significant promise for space amplification reduction, yet existing GC strategies often fall short in optimization performance, lacking thorough consideration of workload characteristics. Additionally, current KV-separated LSM-trees also ignore the adverse effect of the space amplification in the index LSM-tree. In this paper, we systematically analyze the sources of space amplification of KV-separated LSM-trees and introduce Scavenger, which achieves a better trade-off between performance and space amplification. Scavenger initially proposes an I/O-efficient garbage collection scheme to reduce I/O overhead and incorporates a space-aware compaction strategy based on compensated size to minimize the space amplification of index LSM-trees. Extensive experiments show that Scavenger significantly improves write performance and achieves lower space amplification than other KV-separated LSM-trees (including BlobDB, Titan, and TerarkDB).
\end{abstract}


\section{Introduction}
Persistent key-value store, a fundamental component of modern storage systems, provides storage services for various data-sensitive applications such as distributed databases \cite{cao2020characterizing, cao2020polardb, huang2020tidb, chen2022bytehtap}, file system metadata management \cite{harter2014analysis, aghayev2019file}, and stream processing frameworks \cite{chen2016realtime, carbone2015apache}. With advances in digital technology, the continuous growth of data underscores the importance of persistent key-value storage in modern data centers \cite{luo2020lsm}. LSM-tree-based key-value stores have gained widespread adoption in the industry due to their exceptional performance among various persistent key-value storage engines \cite{cassandra, hbase}.

The LSM-tree, leveraging the superior performance of persistent devices for sequential over random data writing \cite{arpaci2018operating}, transforms random writes into sequential ones using an in-memory write buffer, boosting data processing efficiency \cite{o1996log}. It also enables efficient data querying, including point and range queries, by maintaining a hierarchical structure with a globally ordered data layout at each level. However, the LSM-tree's background compaction operations, ensuring multi-level structure maintenance and data ordering, involve frequent storage device reads and writes. This leads to significant read and write amplification \cite{lu2017wisckey, raju2017pebblesdb, wu2015lsm}, attracting notable criticism.

To alleviate the I/O overhead from compaction, the key-value (KV) separation scheme proposed by WiscKey has been extensively adopted in subsequent solutions, including BlobDB\cite{blobdb}, Titan\cite{titan} and TerakDB\cite{terarkdb}. KV-separated LSM-trees reduce LSM-tree size by storing values in a separate log, retaining only index data within the LSM-tree. This approach minimizes the volume of data involved in compaction, reducing write amplification \cite{chan2018hashkv, shennovkv, li2021differentiated}. However, this significantly increases space cost, which can be unbearable in cloud computing environments \cite{zhang2022sa}. The KV-separated LSM-trees alter the effective control over space amplification inherent in vanilla LSM-trees due to the lack of awareness of value size. The prevalent LSM-tree solution, RocksDB \cite{rocksdb}, has shifted its emphasis from striving for ultimate performance to optimizing efficiency \cite{dong2017optimizing, dong2021evolution}, focusing on reducing cost-related space amplification. This shift in development priorities is driven by the understanding that reducing space amplification allows for deploying more instances on the same hardware resources, optimizing the utilization of bandwidth provided by SSDs\cite{dong2017optimizing}. Although RocksDB effectively manages space amplification, its foreground performance is considerably lower than that of KV-separated LSM-trees. For example, under a workload with a fixed value length of 16KB (Fixed-16K), RocksDB's performance is only 0.18x BlobDB's \cite{blobdb}. To maximize performance while maintaining controllable storage space costs, we pursued a solution that strikes a balance between space cost and foreground performance.

In KV-separated LSM-trees, the garbage collection (GC) operation reclaims storage space but also incurs substantial I/O overhead, thereby degrading system performance \cite{chan2018hashkv}. Recent studies have attempted to optimize GC operations for foreground performance improvement\cite{chan2018hashkv, li2021differentiated, shennovkv}. However, these efforts often overlook space reclamation efficiency, leading to exacerbated space amplification due to delayed reclamation \cite{blobdb, li2021differentiated}. For instance, BlobDB employs a compaction-triggered GC policy to mitigate GC-Lookup and GC-write overhead \cite{blobdb, li2021differentiated}. However, this approach leads to severe space amplification because value files must exhaust their data through compaction before being reclaimed. Consequently, under a Fixed-4K workload, it can result in a 3.4x space amplification, significantly higher than other KV-separated LSM-trees. Additionally, current research neglects another primary source of space amplification in KV-separated LSM-trees: space amplification in the index LSM-tree, which often dictates the proportion of hidden garbage in the value data. Existing solutions fail to detect this hidden garbage until the corresponding keys merge in the index LSM-tree, thereby delaying GC scheduling and space reclamation, and leading to severe space amplification. For example, under a Fixed-8K workload, the index LSM-tree's space amplification accounts for 48.8\% of total space amplification. Moreover, diverse workload characteristics render GC operations and space reclamation more intricate. 
In summary, the significant I/O overhead in GC, severe space amplification in the index LSM-tree, and adaptability to varying workloads pose challenges to achieving a space-time balance in KV-separated LSM-trees.

To address these challenges, we propose Scavenger, which provides better space-time trade-offs for KV-separated LSM-trees. Different from previous approaches, we are pioneers in investigating the space amplification problem of KV-separated LSM-trees from a cost perspective. We first conduct a systematic analysis of the fundamental causes of space amplification in KV-separated LSM-trees and then improve efficiency by controlling space overhead while providing high performance. To this end, we optimize GC operations and the compaction strategy to accelerate space reclamation without compromising foreground performance. As a result, Scavenger achieves an improved balance between performance and space utilization by incorporating the above optimizations.

Our principal contributions are summarized as follows:
\begin{itemize}
    \item We quantify that \textbf{space amplification in KV-separated LSM-trees} not only encompasses exposed garbage in the value data, but also includes space amplification in the index LSM-tree, thereby correlating two critical operations, \textbf{garbage collection (GC)} in the value store and \textbf{compaction} in the index LSM-tree.
\end{itemize}
\begin{itemize}
    \item We propose an \textbf{I/O-efficient garbage collection scheme} accelerating three critical GC steps: read, GC-Lookup, and write. It minimizes I/O overhead in GC operations to speed up GC execution, accelerating storage space recycling to reduce space amplification.
\end{itemize}
\begin{itemize}
    \item We propose a \textbf{space-aware compaction strategy based on compensated size}, which converts a separated LSM-tree into a non-separated one. This strategy enables leveled compaction to effectively control space amplification in the index LSM-tree and improves the efficiency of GC.
\end{itemize}
\begin{itemize}
    \item Evaluation results demonstrate that our solution, Scavenger, significantly {\bf improves write performance and accomplishes lower space amplification} than other KV-separated LSM-trees, thereby achieving a better trade-off between performance and space amplification.
\end{itemize}

The organization of this paper is as follows. In Section \uppercase\expandafter{\romannumeral2}, we analyze the trade-offs of existing solutions and quantitatively examine the underlying sources of space amplification for KV-separated LSM-trees. Section \uppercase\expandafter{\romannumeral3} delves into Scavenger's design and discusses the applicability and availability of our solution. The evaluation results are presented in Section \uppercase\expandafter{\romannumeral4}, followed by a discussion of related work in Section \uppercase\expandafter{\romannumeral5}. The paper concludes in Section \uppercase\expandafter{\romannumeral6}.

\section{Preliminaries}

\subsection{Log-Structured Merge-tree}
\begin{figure}[tbp]
    \setlength{\abovecaptionskip}{5pt}
    \setlength{\belowcaptionskip}{-0.6cm}
    \centering
    \subfloat[Vanilla LSM-tree]{
            \includegraphics[width=.42\columnwidth]{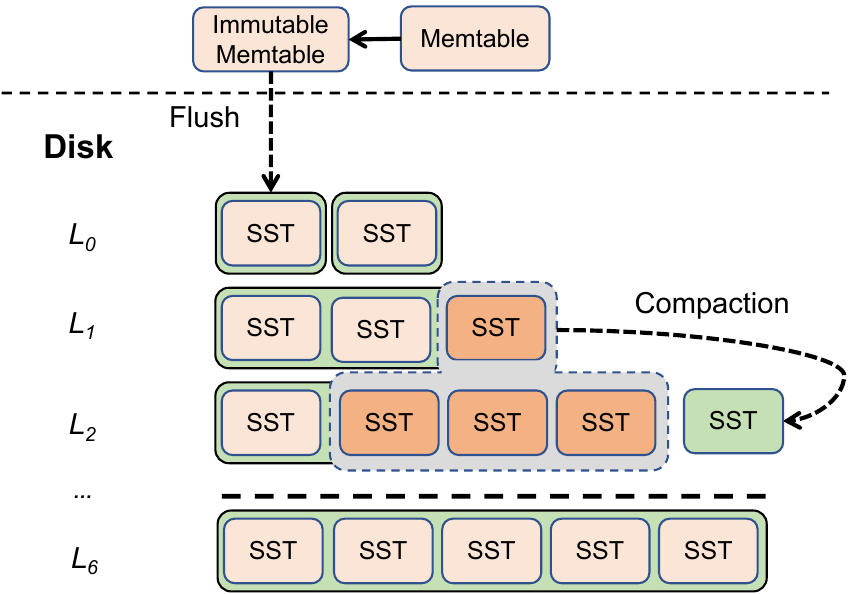}
            \label{fig:bg1-lsm}
        }
    \subfloat[Key-Value Separation: TerarkDB]{
            \includegraphics[width=.53\columnwidth]{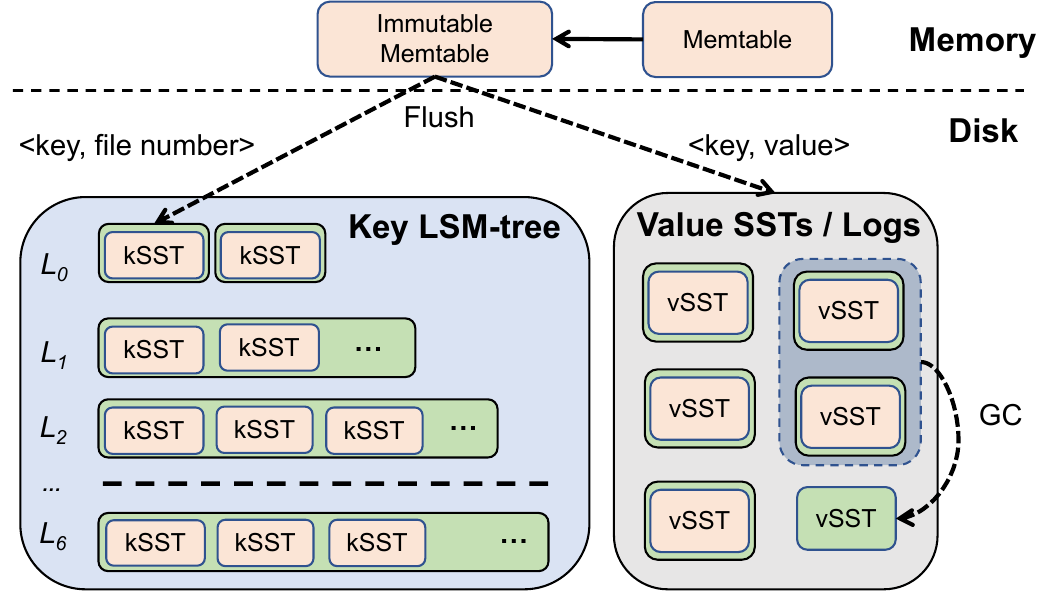}
            \label{fig:bg2-tdb}
        }
    \label{fig:bg1}
    \caption{Overview of LSM-tree and TerarkDB.}
\end{figure}
Vanilla LSM-tree \cite{o1996log} typically employs a hierarchical structure that consists of Memtables in memory and sorted string tables (SSTable, simplified as SST) on disk, as depicted in Figure~\subref*{fig:bg1-lsm}. Memtables are primarily used for buffering data writes and initial sorting, while SSTables, stored on storage devices, maintain a multi-level structure to ensure query performance. Within the LSM-tree, the $L_0$ level consists of multiple SSTables, facilitating flush operations from memory, and the $L_1$ to $L_N$ levels are organized in a sorted manner, with multiple SSTables forming a single level for efficient query.

\textbf{Write process}. Initially, data requests are recorded in the Write-Ahead Log (WAL) to ensure crash consistency. After recording, the data is temporarily stored in the memory Memtable. Once the Memtable is full, the state of the Memtable becomes immutable, and the Immutable Memtable is flushed to $L_0$ as an SSTable (SST). When the level $L_i$ becomes full, the LSM-tree schedules the background compaction thread to select a set of overlapping SSTables from two adjacent levels $L_i$ and $L_{i+1}$, merges key-value pairs from these SSTables, and creates new SSTables that are written to the $L_{i+1}$ level. Compaction involves multiple I/Os of storage devices, leading to significant write amplification.

\textbf{Read process}. Query operations are categorized into point and range queries. For point queries, the LSM-tree traverses each component from memory to storage, including Memtables and SSTables. If a component potentially encompassing the key-value pair being queried is identified, a binary search is conducted within it. When querying an SSTable, the LSM-tree employs index and filter blocks to retrieve data. Range queries commence by locating the starting key in each component, and then reading key-value pairs within the specified range by moving the iterators of the components. Ultimately, a collection of key-value pairs will be returned to the users.

\subsection{Key-Value Separation: A Case Study of TerarkDB}
To address the significant write amplification caused by compaction operations in vanilla LSM-trees, WiscKey \cite{lu2017wisckey} innovatively proposed the KV-separated LSM-tree. Following that, as a recognized LSM-tree optimization technique, KV separation has been widely applied in the industry, such as BlobDB \cite{blobdb}, BadgerDB \cite{badger}, TitanDB \cite{titan}, and TerarkDB \cite{terarkdb}. Subsequently, many studies on KV-separated LSM-trees focus on improving foreground performance \cite{chan2018hashkv, shennovkv, li2021differentiated, tang2022fencekv}. 

We take TerarkDB, a high-performance and widely used KVS developed by ByteDance, as an example of KV-separated LSM-trees. It serves as the underlying storage engine for various systems within ByteDance, such as the distributed OLTP database, ByteNDB \cite{chen2022bytehtap}, and stream processing frameworks, Flink \cite{flink}. Figure~\subref*{fig:bg2-tdb} depicts TerarkDB's architecture. TerarkDB stores keys and references to key-value pairs ($<$$key, file \ number$$>$) in vanilla LSM-tree's SSTables (denoted as kSST) while storing values in multiple SSTables (denoted as vSST). Unlike the unordered value log in WiscKey, TerarkDB employs SSTables to store values in order, improving the performance of range queries. Similar to WiscKey, TerarkDB also schedules background GC threads to reclaim the storage space occupied by obsoleted vSSTs.

Compared to other KV-separated LSM-trees, TerarkDB shows superior foreground performance with its no-writeback GC strategy. Specifically, TerarkDB employs a tuple consisting of the key and associated vSST file number, where the value resides, as the index to store in the LSM-tree. Moreover, TerarkDB maintains the inheritance relationship between the latest vSST file number and the outdated vSST file number after GC instead of updating the address of the valid key-value pairs in the LSM-tree index, which is a widely used method \cite{blobdb, badger, titan}. As a result, TerarkDB can find the most recent vSST file number where the value currently resides and retrieve the corresponding data by accessing the index blocks within vSST. In short, TerarkDB minimizes the impact of GC operations on foreground writes by avoiding index rewriting.

\begin{figure}[tbp]
    \setlength{\abovecaptionskip}{5pt}
    \setlength{\belowcaptionskip}{-0.6cm}
    \centering
    \subfloat[Update throughput]{
            \includegraphics[width=.48\columnwidth]{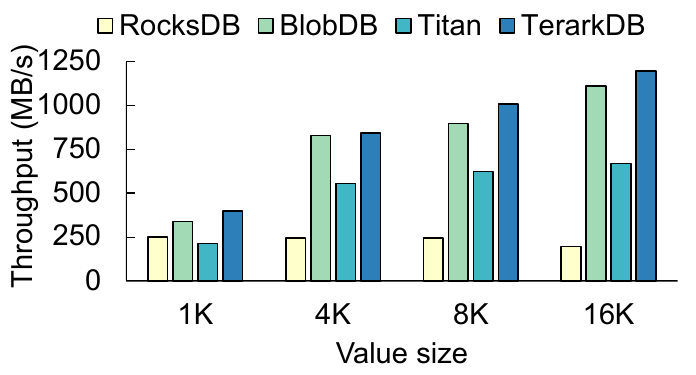}
            \label{fig:mot1-1}
        }
    \subfloat[Space amplification]{
            \includegraphics[width=.48\columnwidth]{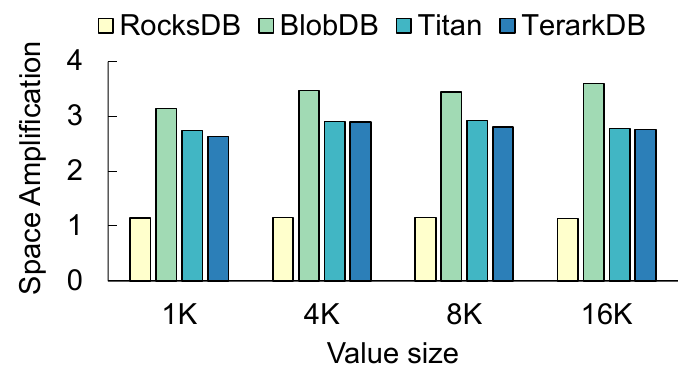}
            \label{fig:mot1-2}
        }
    \caption{Space-time trade-offs of existing solutions.}
    \label{mot1}
\end{figure}

\subsection{Motivation}
\subsubsection{Trade-offs of Existing Solutions}
The performance improvement of KV-separated LSM-trees over vanilla LSM-trees is attributed to their effective GC operations, which reclaim space with reduced read and write amplification \cite{chan2018hashkv, li2021differentiated}. However, this benefit is often accompanied by the risk of increased space amplification. As depicted in Figure \ref{mot1}, we evaluated the update performance and space amplification of various advanced KV-separated LSM-trees, including BlobDB, Titan, TerarkDB, and the widely used LSM-tree key-value store, RocksDB. Compared to RocksDB, KV-separated LSM-trees improve update throughput by 2.57x-4.16x for the 8KB workload, but this comes at the cost of 2.42x-2.97x more space usage. We set the garbage ratio threshold for triggering GCs at 20\%, which means the expected space amplification is 1/(1-20\%)=1.25. However, KV-separated LSM-trees experience excessive space amplification that exceeds this expectation, overlooking actual disk capacity and total cost of ownership (TCO). Simultaneously, the varying characteristics of workloads pose challenges to the foreground performance and space reclamation of KV-separated LSM-trees.

Significant space amplification of existing KV-separated LSM-trees leads to high TCO, making it challenging to scale like a single-disk multi-instance. We aim to achieve improved space-time trade-offs across diverse workloads within KV-separated LSM-trees. To this end, it is necessary to clarify the fundamental sources of space amplification. Therefore, we have selected TerarkDB and Titan, two widely-used KV-separated LSM-trees in the industry \cite{li2021differentiated, li2021elastic}, to serve as our case studies.
We aim to investigate the root cause of the space amplification issue and to explore more effective solutions.

\subsubsection{GC Operations Analysis}
GC efficiency is a key determinant of space amplification magnitude\cite{xanthakis2021parallax}. To further understand the sources of GC overhead under diverse workloads, we dissect GC's workflow and break down the latency of GC operations in TerarkDB and Titan.

\textbf{Workflow of GC operations}. In traditional KV-separated LSM-trees like WiscKey and Titan, GC operations involve four steps: reading eligible data (denoted as \textbf{Read}), verifying its validity (denoted as \textbf{GC-Lookup}), writing valid data to a new value log (denoted as \textbf{Write}), and rewriting the latest index to the index LSM-tree (denoted as \textbf{Write-Index}). Conversely, TerarkDB stores file numbers instead of addresses in the index LSM tree and maintains file number inheritance relationships between outdated and latest vSSTs, thus avoiding index updates (\textbf{Write-Index}). As a result, the GC process in TerarkDB consists of three steps: \textbf{Read}, \textbf{GC-Lookup}, and \textbf{Write}. For \textbf{Read}, TerarkDB traverses the entire vSST file to read key-value pairs from candidate files. For \textbf{GC-Lookup}, TerarkDB queries the index LSM-tree to obtain the vSST file number, comparing it with the current GC vSST file number. If the numbers mismatch, the key-value pair is deemed garbage and should be discarded. Otherwise, the key-value pair is considered valid and rewritten to the new vSST as a \textbf{Write}.

\textbf{GC bottlenecks under various workloads}. We conducted a latency breakdown of critical steps in the GC processes of TerarkDB and Titan to identify performance bottlenecks. The evaluations were initiated by loading 100GB of unique data and updating 300GB to trigger frequent GC operations. We evaluated the latency percentage for each step, obtained by $\frac{latency_{step}}{latency_{GC}}$, and the average latencies for both fixed-length and variable-length workloads, respectively. Fixed-length workloads encompass five sets, with different value sizes ranging from 1KB to 16KB. Variable-length workloads include mixed workloads with a 1:1 ratio of large to small values and workloads where value sizes follow a generalized Pareto distribution \cite{hosking1987parameter, rocksdb-trace}. The average sizes for the Mixed and Pareto workloads are 8K and 1K, respectively. Notably, the Mixed workload, a typical pattern in ByteDance's internal OLTP database, exhibits a 1:1 ratio of large (16KB) to small (uniformly distributed from 100 to 512B) values. Large values primarily stem from updates to the original data pages of DB, while small values come from incremental user writes.

\begin{figure}[tbp]
    \setlength{\abovecaptionskip}{5pt}
    \setlength{\belowcaptionskip}{-0.6cm}
    \centering
    \includegraphics[width=0.47\textwidth]{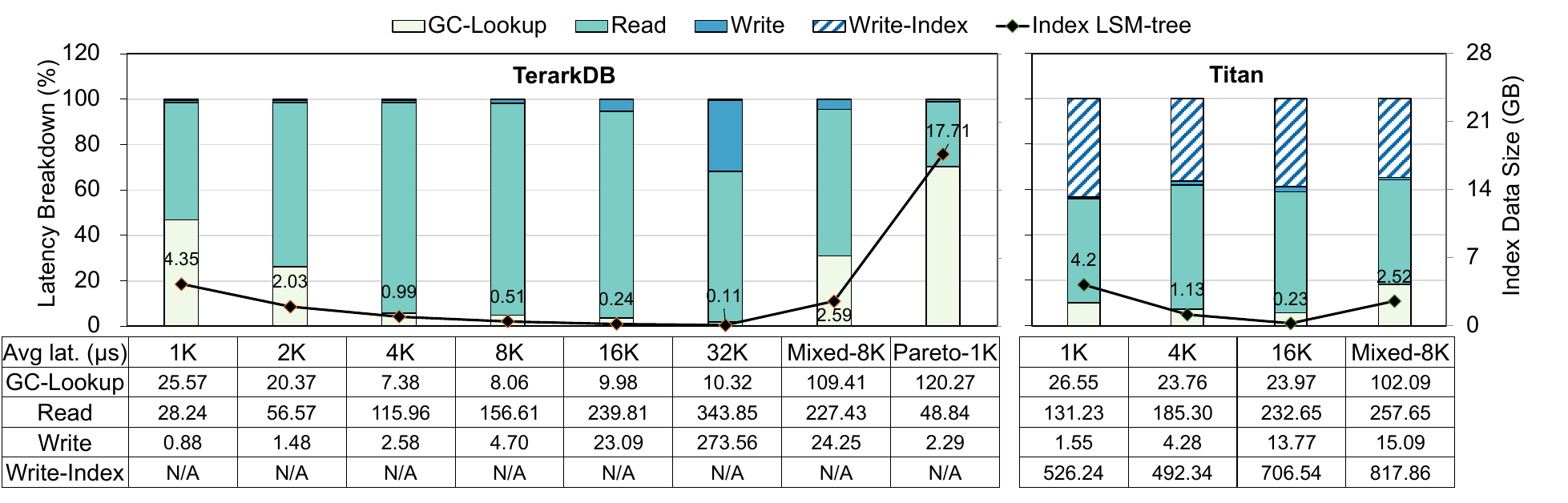}
    \caption{GC latency breakdown of TerarkDB and Titan.}
    \label{fig:mot4-gc-lat}
\end{figure}

Figure~\ref{fig:mot4-gc-lat} shows the latency breakdown of each step under different workloads, indicating that GC operations' performance is highly workload-dependent. Specifically, \textbf{Read} has the highest latency percentage for most workloads, exceeding 50\% for all workloads except for Pareto-1K. In fixed-length workloads, \textbf{Read} and \textbf{Write} proportions rise with value size, while the \textbf{GC-Lookup} proportion inversely grows with decreasing value size. Notably, the overhead of \textbf{GC-Lookup} is significant for variable-length workloads, especially for Pareto-1K. In Titan, \textbf{Write-Index} incurs substantial overhead, about 38\% of GC. If \textbf{Write-Index} is removed as in TerarkDB, \textbf{Read} and \textbf{GC-Lookup} overhead percentages increase by 25\% and 10\% for Mixed-8K, becoming significant performance bottlenecks. Moreover, Titan's GC \textbf{Read} latency is much higher than TerarkDB's, as it is not accelerated using cache.

A deeper analysis reveals that the latency of \textbf{GC-Lookup} is primarily influenced by the size of the index LSM-tree and the mixed storage layout for index and small values. The bottom table in Figure~\ref{fig:mot4-gc-lat} illustrates that the average latencies of \textbf{Read} and \textbf{Write} rise with value size, whereas the average latency of \textbf{GC-Lookup} consistently increases with the size of the index LSM-tree. Specifically, index LSM-tree sizes for variable-length workloads exceed those for fixed-length workloads, particularly for Pareto-1K, leading to a higher latency percentage. Notably, despite the index LSM-tree size under the Mixed-8K workload being comparable to Fixed-2K, the average latency of \textbf{GC-Lookup} escalates by 5.4x. Further analysis reveals a 22\% decrease in the cache hit ratio under Mixed-8K compared to Fixed-2K, attributable to the mixed storage layout for index and small values.

\subsection{Another source of Space Amplification: Index LSM-tree}
\subsubsection{Space Amplification Analysis}
Although the efficiency of GC affects space amplification, another frequently overlooked primary source in KV-separated LSM-trees is within the index LSM-tree. In light of this, we conducted a systematic analysis to investigate space amplification thoroughly.

\begin{figure}[tp]
    \setlength{\abovecaptionskip}{5pt}
    \setlength{\belowcaptionskip}{-0.6cm}
    \centering
    \includegraphics[width=0.41\textwidth]{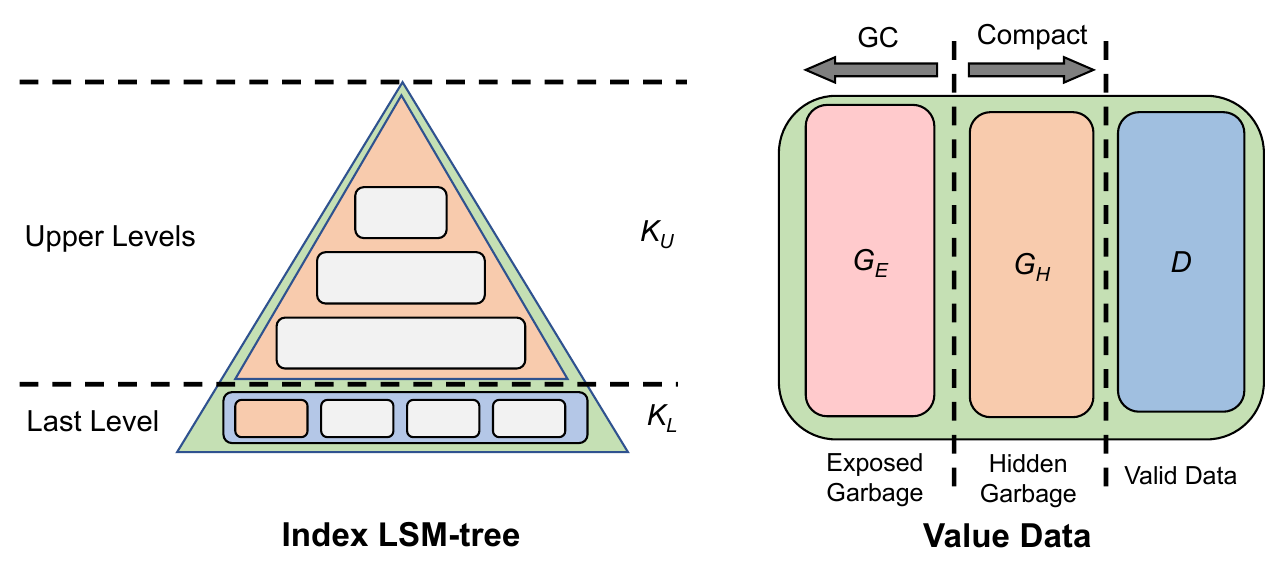}
    \caption{Space amplification analysis.}
    \label{fig:mot5-space-model}
\end{figure}

For vanilla LSM-trees, we first quantify space amplification by analyzing its structure. Upon enabling Dynamic Capacity Adaptation (DCA) \cite{dynamic-level, dong2017optimizing} and reaching a steady state, the size of the LSM-tree's last level, given the fixed and large inter-level ratio (default 10), provides a reasonable estimate of user data size \cite{dynamic-level, dayan2022spooky}. The total KVS size divided by the last level size gives a reasonable space amplification estimate \cite{dynamic-level}. As shown in Figure \ref{fig:mot5-space-model}, we denote the size of the last level as $K_L$ and the total size of the upper levels as $K_U$ so that we can calculate the space amplification as 
\begin{equation}
    S_{Index} \approx \frac{K_U+K_L}{K_L}=\frac{K_U}{K_L}+1
    \label{eq:lsm_sa}
\end{equation}
Assuming that the inter-level ratio of the LSM-tree is 10, the LSM-tree can achieve 1.11x space amplification. We use this estimate to simplify the space amplification model, which helps to understand the source of space amplification.

For KV-separated LSM-trees, considering that the size of value data typically exceeds that of the index LSM-tree, we can estimate the overall space amplification of the KVS from the space amplification of the value data. The space amplification of value data is primarily influenced by the proportion of garbage data generated from user writes/deletes. However, the generation of garbage data is delayed due to the out-of-place writing mechanism, which delays the execution of data merging and deletion. Thus, we can divide the value data into the following components:
\begin{itemize}
    \item \textbf{Valid Data $D$} represents the value data associated with the last level of the index LSM-tree, denoting the total size of the unique dataset.
    \item \textbf{Hidden Garbage $G_H$} corresponds to the value data associated with keys in the last level of the index LSM-tree that are still awaiting merge. Even though it is obsolete, the system cannot yet recognize it as garbage, as the merge has yet to occur.
    \item \textbf{Exposed Garbage $G_E$} refers to the value data of the keys that were merged during the index LSM-tree's compaction. It is incorporated into the garbage ratio computed by the system and triggers subsequent GC operations for space reclamation.
\end{itemize}

Given that the LSM tree inherently tolerates the space amplification of upper-level data, this leads to the hiding of garbage data. Therefore, the \textbf{Hidden Garbage} to \textbf{Valid Data} ratio can be approximated by the proportion between upper levels and last-level data in the index LSM-tree, as defined by
\begin{equation}
\frac{G_H}{D} \approx
\frac{K_U}{K_L}
\label{eq:ratio}
\end{equation}

Based on the analysis above, the space amplification in KV-separated LSM-trees ($S_{value}$) can be expressed as
\begin{equation*}
S_{value} = \frac{G_E+G_H+D}{D}=\frac{G_E}{D}+(\frac{G_H}{D}+1) \approx \frac{G_E}{D}+(\frac{K_U}{K_L}+1)
\label{eq:sa}
\end{equation*}
\begin{equation}
S_{value} \approx \frac{Exposed \ Garbage}{Valid \ Data}+S_{Index}
\label{eq:sa2}
\end{equation}
Specifically, the size of \textbf{Valid Data} $D$ is generally determined by the size of the dataset, therefore \textbf{Exposed Garbage $G_E$} and \textbf{Hidden Garbage $G_H$} constitute the main contributors to space amplification. Compaction decreases the space amplification of the index LSM-tree, thus transforming hidden garbage into exposed garbage. Meanwhile, GC assists in reducing the exposed garbage, further lowering the space amplification.

\begin{figure}[tbp]
    \setlength{\abovecaptionskip}{5pt}
    \setlength{\belowcaptionskip}{-0.6cm}
    \centering
    \subfloat[Space amplification of index]{
            \includegraphics[width=.47\columnwidth]{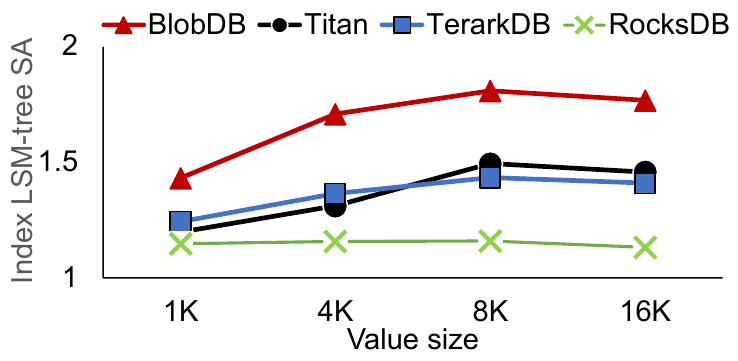}
            \label{fig:mot2-1}
        }
    \subfloat[Exposed garbage of value]{
            \includegraphics[width=.47\columnwidth]{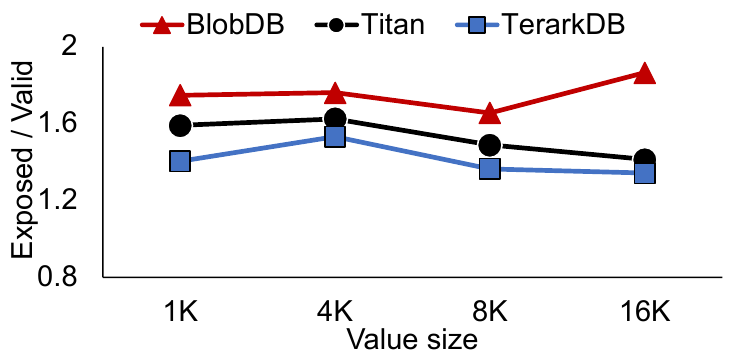}
            \label{fig:mot2-2}
        }
    \caption{Two sources of space amplification.}
    \label{mot2}
\end{figure}

We investigated the direct cause of space amplification for KV-separated LSM-trees by evaluating the space amplification of the index LSM-tree $S_{Index}$ and the \textbf{Exposed/Valid} ratio in the value store. Figure \ref{mot2} shows $S_{Index}$ surpassing 1.11x (the ideal space amplification of RocksDB) and \textbf{Exposed/Valid} ratio robustly exceeds 0.25 (the ideal ratio of no hidden garbage corresponds to the 20\% garbage rate setting, which is $\frac{20\%}{80\%} = 0.25$). Using the space amplification of TerarkDB under the 8K workload as an example, we discovered that the space amplification of the index LSM-tree and the exposed garbage in the value data accounted for 51.2\% and 48.8\% of the total space amplification, respectively. Other KV-separated LSM-trees demonstrate similar behaviors under diverse workloads, albeit with varying proportions.

\subsubsection{Compaction of KV-separated LSM-trees}
We further investigated the reasons for the significant space amplification of index LSM-trees by analyzing the corresponding LSM-tree shape and file sizes. For Fixed-8K workload, KV-separated TerarkDB flushes small kSST files to the $L_0$ level, with the average file size being only about 211K, much less than the 64M in a vanilla LSM-tree. The smaller file size typically makes reaching the level-size-based compaction trigger threshold more challenging, thus delaying compaction. This results in the number of compactions after separation being only 27.8\% of that of a vanilla LSM-tree under Fixed-8K. From a cost perspective, it is crucial for LSM-trees to initiate compaction promptly to ensure the convergence of space amplification. Delayed compaction always results in persistent and substantial space amplification, as shown in Figure \subref*{fig:mot2-1}. This is due to the substantial amount of upper-level data that cannot be merged with lower-level data.

To make matters worse, different workloads result in different file sizes in the index LSM-tree, and a static compaction trigger threshold proves suboptimal for accommodating such diversity in workloads. Therefore, it is an urgent problem to figure out how to schedule compaction more efficiently to maintain the ideal space amplification of the index LSM-tree.

\section{Design}
We introduce Scavenger, a novel KV-separated LSM-tree KV store designed to achieve better space-time trade-offs.

\subsection{System Overview}
\begin{figure}[tbp]
    \setlength{\abovecaptionskip}{5pt}
    \setlength{\belowcaptionskip}{-0.6cm}
    \centering
    \includegraphics[width=0.47\textwidth]{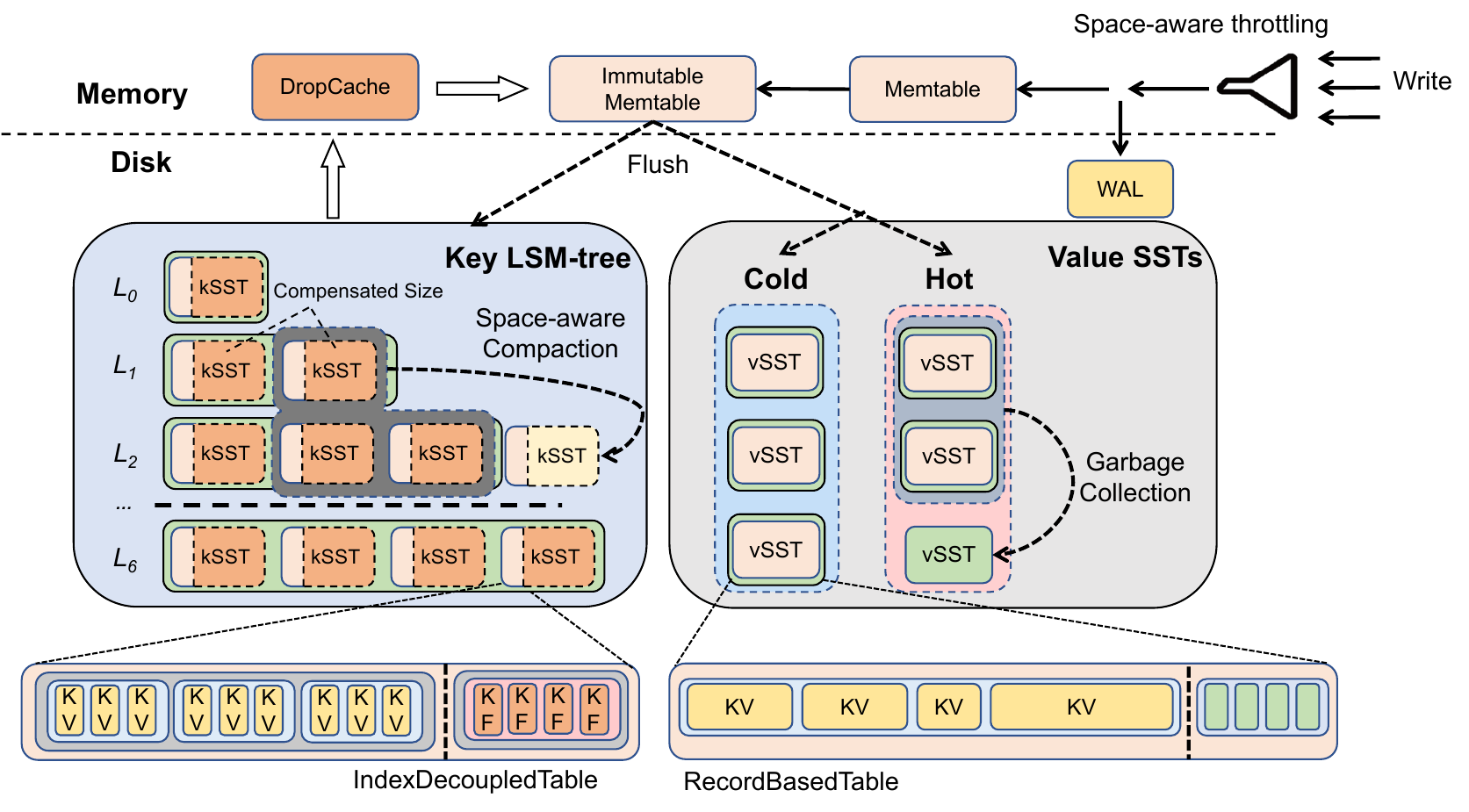}
    \caption{Overview of Scavenger.}
    \label{fig:design1-all}
\end{figure}

Similar to the most KV-separated LSM-trees, Scavenger utilizes a hierarchical data management architecture. It integrates standard LSM-tree components, including Memtables, Immutable Memtables, and Write-Ahead Logs, as depicted in Figure~\ref{fig:design1-all}. Concurrently, Scavenger separates data into an index LSM-tree for key indexing and several value tables for values.

Distinct from other solutions, Scavenger provides better space-time trade-offs for KV-separated LSM-trees by refining GC and compaction procedures. For GC operations, Scavenger incorporates three novel structures. It introduces \textbf{RecordBasedTable}, a GC-compatible structure designed to enhance GC read operations for value SST. Scavenger utilizes \textbf{IndexDecoupledTable} to separate small key-value and index data, thereby optimizing GC-Lookup operations for key SST. Additionally, Scavenger introduces \textbf{DropCache} to record the keys dropped during compaction and implements a lightweight, hotness-aware write strategy for hot-cold data separation. For compaction operations, Scavenger adopts a space-aware compaction strategy based on compensated size to minimize space amplification in the index LSM-tree. Finally, Scavenger applies space-aware throttling before processing write requests to control space usage.

\subsection{I/O-Efficient Garbage Collection}
Our previous analysis identified three primary sources of overhead in the garbage collection (GC) process: \textbf{Read}, \textbf{GC-Lookup}, and \textbf{Write}. To address these issues, we propose an I/O-efficient garbage collection scheme. This scheme has been specifically designed to minimize I/O costs during the GC process by optimizing these three critical operations.

\subsubsection{Lazy Read of the GC}
Our analysis in Section \uppercase\expandafter{\romannumeral2}-C identifies GC Read as the primary latency bottleneck in the garbage collection (GC) procedure for fixed-length, large-value workloads. As value size grows, GC Read experiences escalating latencies, thus dominating the cost of value access. From a holistic perspective of GC operations, the volume of data processed during GC Read typically surpasses that of GC Write. This discrepancy primarily arises because the data's validity can only be verified after reading, and then only the valid entries are rewritten. Consequently, the GC Read operation accesses a substantial amount of unnecessary garbage, which will be discarded during GC. Therefore, reducing access to this garbage data is paramount to mitigating GC Read overhead.

An intuitive approach would be to read only data that is deemed valid. However, it is often impractical to store sufficient validity information as prior knowledge within a limited overhead for large-scale persistent key-value stores. A more feasible strategy is to minimize the amount of garbage data read, i.e., to minimize the data read for validity checks and then read the remaining valid data once the validity check has passed. Validation (GC-Lookup) in KV-separated LSM-trees typically involves executing a point query operation on the index LSM-tree for a specific key. Therefore, we can validate the entire vSST by examining only the contained keys. After obtaining the validity information, we validate key-value pairs and avoid unnecessary value reading for obsolete pairs.

\begin{figure}[tbp]
    \setlength{\abovecaptionskip}{5pt}
    \setlength{\belowcaptionskip}{-0.6cm}
    \centering
    \includegraphics[width=0.40\textwidth]{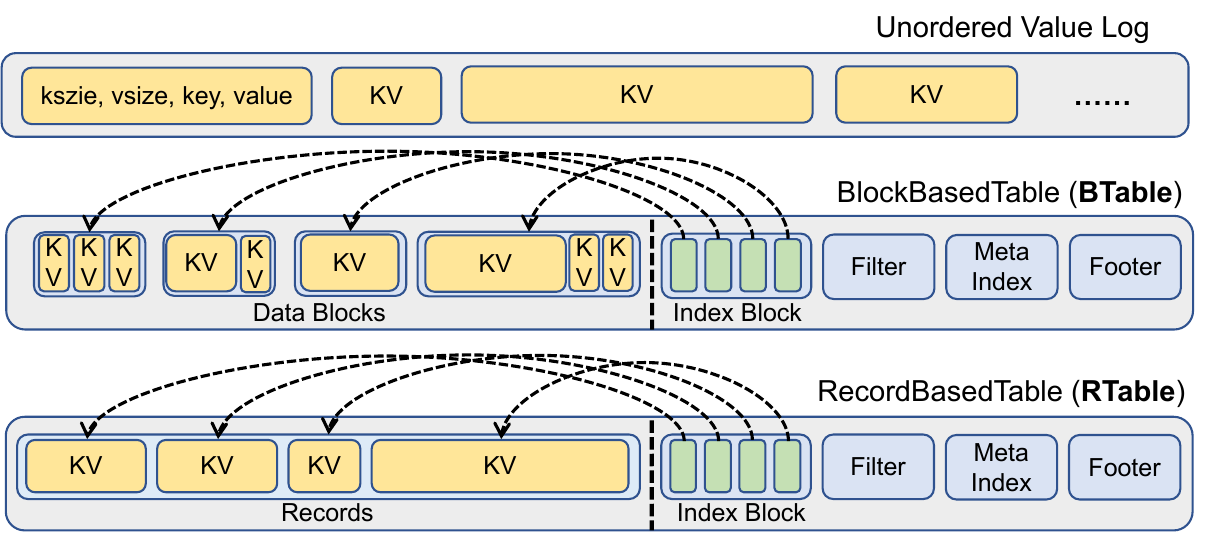}
    \caption{Storage layout of RTable.}
    \label{fig:design2-read}
\end{figure}

\textbf{Storage layout of vSST.}
Redesigning the vSST storage layout is essential for efficient key retrieval. Conventional methods, such as the \textbf{unordered value log} of WiscKey, store the $<$$key\_size, value\_size, key, value$$>$ tuple as a log record, as shown in Figure~\ref{fig:design2-read}. A fully randomized storage layout complicates accessing all keys in the log. Industrial schemes like TerarkDB adopt a structure similar to RocksDB's SSTable, known as \textbf{BlockBasedTable} (\textbf{BTable}). While the BTable's internal index block facilitates retrieval, it only points to data blocks, which may contain multiple entries. In essence, BTable maintains a sparse index for key-value pairs, which hinders the rapid retrieval of keys with minimal I/O.

To efficiently retrieve all keys, Scavenger introduces a novel table structure, \textbf{RecordBasedTable}, abbreviated as \textbf{RTable}. This structure is specifically designed to store values in KV-separated LSM-trees. Like BTable, RTable maintains data orderly and leverages existing metadata (Footer, Meta Index, and Filter) to enhance data retrieval. However, unlike traditional BTable, RTable organizes values as records (key-value pairs) and creates a $<$$key, offset$$>$ tuple as the index for each record in the index block. Essentially, RTable creates a dense index for key-value pairs, allowing GC read operations to directly access the RTable's index blocks to retrieve all keys without accessing values. Simultaneously, foreground read operations can retrieve value addresses directly from the index block and read values without in-block retrieval efficiently.

It is worth mentioning that the dense index of the RTable slightly increases the size of the index. For the Pareto-1K workload, RTable introduces about a 2\% extra space overhead, still considerably smaller than the separated value data. Furthermore, setting a larger KV separation threshold can help reduce the storage space. Regarding I/O overhead, we employ a partitioned index \cite{part-index} to reduce the I/O of foreground read operations, thereby accessing only necessary index blocks.

\begin{figure}[tbp]
    \setlength{\abovecaptionskip}{5pt}
    \setlength{\belowcaptionskip}{-0.6cm}
    \centering
    \includegraphics[width=0.37\textwidth]{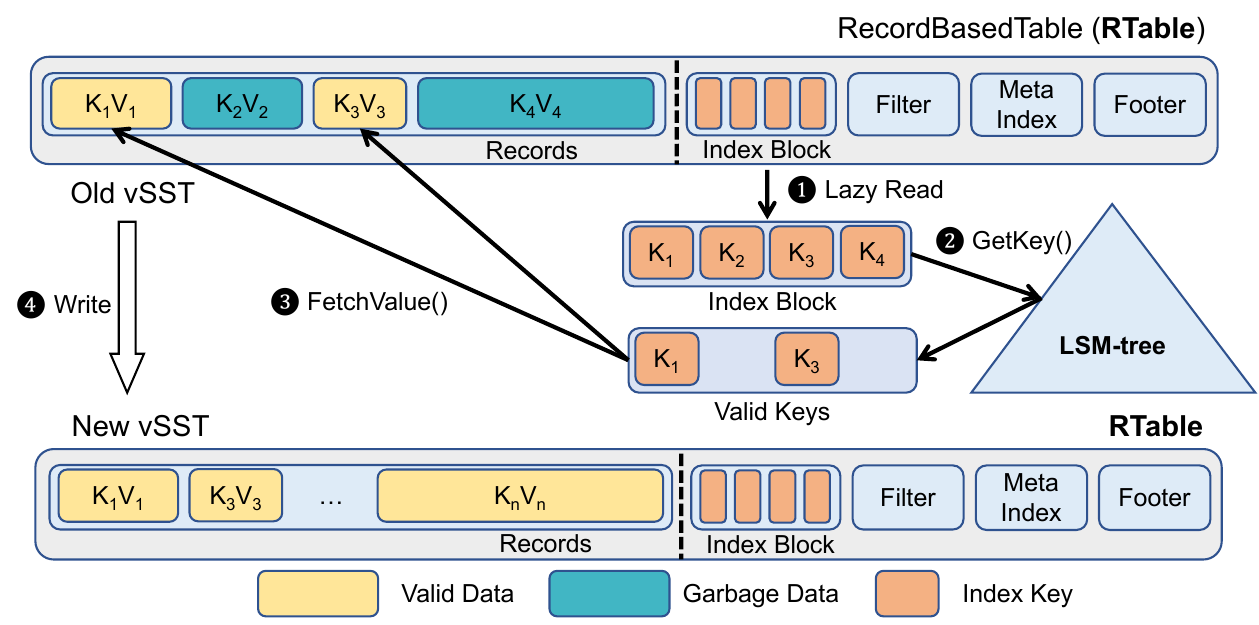}
    \caption{Lazy Read process of GC.}
    \label{fig:design3-read}
\end{figure}

\textbf{Lazy Read}. 
RTable enables efficient GC read operations, requiring minimal I/O to access all keys. Initially, the GC thread reads the footer block to get the index block address, and then retrieves the entire index block from that location. The index block, which contains dense indexes for key-value pairs, allows the GC thread to perform validation checks (GC-Lookup) without accessing values. This effectively postpones the actual value read until valid keys are identified. In essence, RTable assists GC read operations to minimize I/O overhead. The GC read process with RTable is referred to as lazy read.

Figure~\ref{fig:design3-read} illustrates the lazy read process of GC. Initially, the GC thread reads the index block of the selected vSST, acquires keys $K_1$-$K_4$, and stores the index block in the cache with high priority. Then, the GC thread invokes the $GetKey()$ function for GC-Lookup to validate keys. Among these, $K_2$ and $K_4$ are identified as garbage, while $K_1$ and $K_3$ are deemed valid. After GC-Lookup, the GC thread retrieves the values $V_1$ and $V_3$ of the valid keys $K_1$ and $K_3$ using the address information stored in the index block. Then, valid key-value pairs $K_1V_1$ and $K_3V_3$ are appended to the record segment of the newly created vSST. Invalid values are not accessed during GC.

\subsubsection{Index-Record Separation}
The lazy read approach significantly reduces GC read overhead, especially in large key-value workloads. However, as discussed in Section \uppercase\expandafter{\romannumeral2}-C, GC-Lookup is another significant source of GC overhead, particularly in fixed-length small key-value workloads and variable-length workloads. Therefore, optimizing GC-Lookup overhead for these workloads is crucial. Notably, the essence of GC-Lookup is the point query operation on the index LSM-tree, allowing cache and index optimization techniques. Our goal is to explore how KV-separated LSM-trees can utilize existing optimization techniques more efficiently and how to tackle challenges unique to KV-separated LSM-trees.

The evaluations in Section \uppercase\expandafter{\romannumeral2}-C show that LSM-tree sizes for fixed-length 1KB and 2KB exceed the pre-allocated cache size of 1GB (1\% of the dataset). This situation becomes worse in variable-length Mix-8K and Pareto-1K workloads. Allocating more memory for caching the index LSM-tree is beneficial, but practical constraints limit significant memory allocation for high-capacity servers. Consequently, it becomes imperative to optimize cache space utilization and enhance GC-Lookup efficiency. We also focus on the issues unique to KV-separated LSM-trees. Unlike traditional point query operations in vanilla LSM-trees, GC-Lookup operation in KV-separated LSM-trees does not require access to the value data of small KV pairs stored in the index LSM-tree. Instead, it simply necessitates checking the indexes of large KV pairs. However, in current KV-separated LSM-tree implementations, small KV data and index data for large KV are stored within the same data block. As a result, the GC-Lookup operation incurs unnecessary data access for small KV pairs. To tackle this challenge, Scavenger introduces a novel table structure, \textbf{IndexDecoupledTable}.

\begin{figure}[tbp]
    \setlength{\abovecaptionskip}{5pt}
    \setlength{\belowcaptionskip}{-0.6cm}
    \centering
    \includegraphics[width=0.38\textwidth]{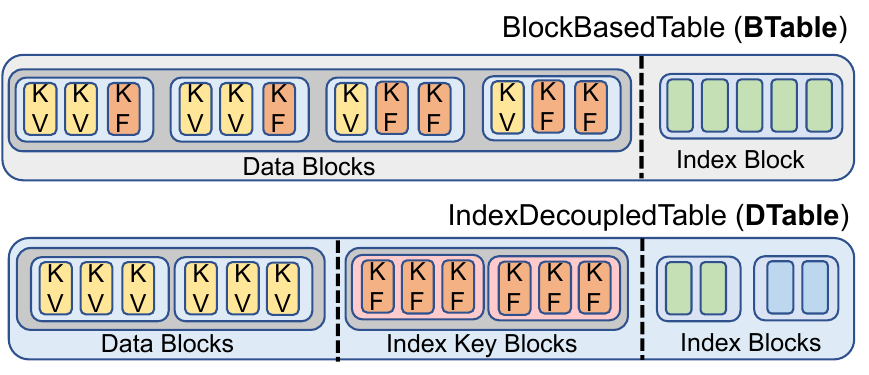}
    \caption{Storage layout of DTable.}
    \label{des5-dtable}
\end{figure}

\textbf{Storage layout of kSST.}
For key SST, existing KV-separated LSM-trees utilize the default BTable structure to store both indexes (represented by \textbf{KF}, $<$$key, address$$>$) and small KV data (also referred to as \textbf{Record}, denoted by \textbf{KV}, $<$$key, value$$>$) within the same file. This combined index and small KV pair storage layout substantially increases GC-Lookup latency as it reads unrelated value data. Specifically, GC-Lookup only needs to examine indexed data (KFs) stored in the LSM-tree. A point query operation accesses at least a data block, but a data block often contains a lot of small-value data (KVs) that GC-Lookup does not need.

Scavenger aims to provide quick access to pure indexes without introducing additional overhead. To this end, Scavenger proposes a cache-efficient table structure, \textbf{IndexDecoupledTable} (\textbf{DTable}), segregating the index entry (KF) and record (KV) in the index LSM-tree. Specifically, as shown in Figure \ref{des5-dtable}, DTable generates independent data blocks and index blocks for index entries (KF) and records (KV), enabling index entries and records to be accessed independently without accessing data of other types. GC-Lookup can directly access index key blocks for rapid validation. Simultaneously, we assign higher cache priority to index key blocks as they store more keys than original data blocks and can serve multiple operations (including foreground queries and GC lookups). This is implemented by placing the index key blocks into the block cache's high-priority queue, which is designed to keep hot data in the cache for longer periods before eviction.

It is worth noting that the design of DTable is fully compatible with BTable, maintaining the same structure in scenarios where all data (large-value) or none of it (small-value) is stored separately. DTable eliminates unnecessary GC-Lookup I/O for variable-length workloads, thereby improving cache efficiency and GC-Lookup speed. For foreground queries, DTable seldom incurs additional I/O from overlapping ranges due to the high-priority caching of index key blocks and active cache replacement during compaction. This design even eliminates one index query I/O for large value data queries, further enhancing the efficiency of DTable.

\begin{figure}[tbp]
    \setlength{\abovecaptionskip}{5pt}
    \setlength{\belowcaptionskip}{-0.6cm}
    \centering
    \includegraphics[width=0.47\textwidth]{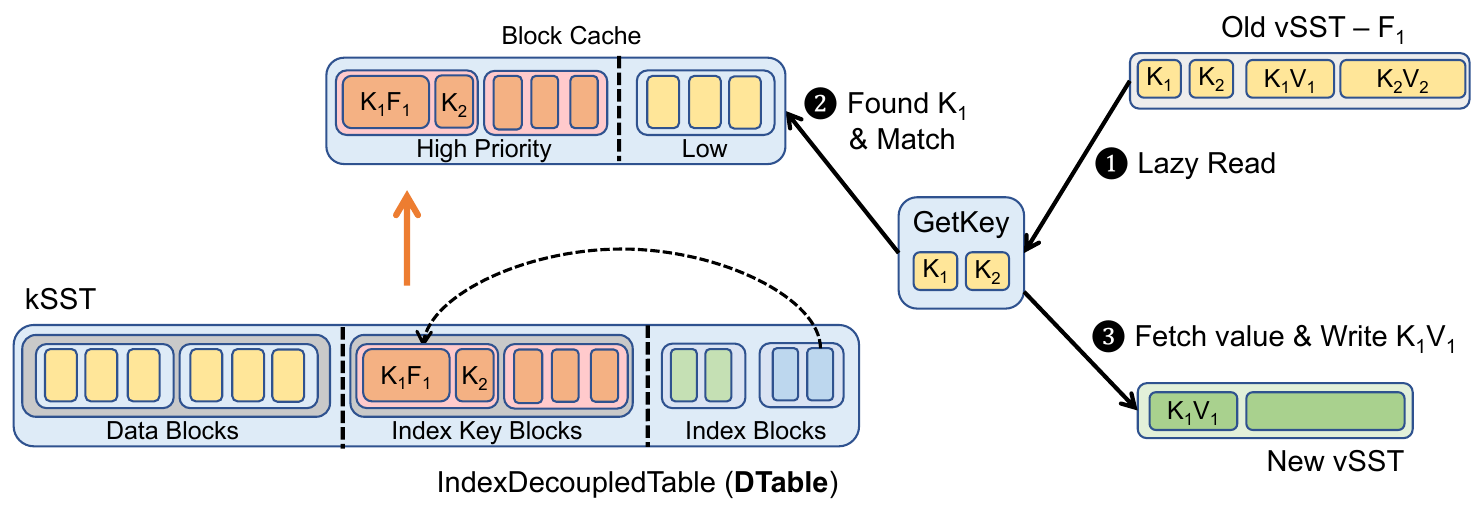}
    \caption{GC-Lookup process.}
    \label{fig:design5-discuss}
\end{figure}

\textbf{GC-Lookup.}
GC-Lookup operations, which are typically characterized by good access locality, involve executing multiple point queries across a continuous key range. As illustrated in Figure \ref{fig:design5-discuss}, the GC thread initially fetches keys $K_1$ and $K_2$. If the validity check on $K_1$ triggers a cache miss, the related index block is loaded from the DTable. Subsequently, the index key block containing $K_1$ is identified and loaded into the block cache with a high priority. GC-Lookup then locates $K_1$ and verifies its validity with the vSST file number. As a result, the corresponding value, $K_1V_1$, is retrieved and rewritten to the new file. In the subsequent GC-Lookup on $K_2$, the query hits the block cache. Since the block cache can serve most GC-Lookup, this significantly accelerates the GC-Lookup process.

\subsubsection{Hotness-aware writing}
Workloads in existing systems exhibit both hot and cold characteristics, in addition to variable length characteristics. Hot and cold workload characteristics significantly impact GC execution efficiency. Specifically, the GC operation reclaims garbage from repetitive data writes, denoted as hot-write data, which includes obsolete data stemming from updates and deletes \cite{zhang2022halsm}. Consequently, a higher proportion of hot-written data implies more garbage in a file, enabling the GC to reclaim more space. To leverage the cold-hot characteristics, Scavenger adopts a hotness-aware write strategy that identifies data hotspots and uses this insight to guide writing operations, including Flush and GC operations.

\textbf{DropCache}.
Scavenger uses an in-memory cache, DropCache, for hotspot detection, which operates as an LRUCache, similar to RocksDB's block cache \cite{rocksdb-bcache}, but records keys. Hot-write data, mainly from user actions like deletions and overwrites, is eventually merged during compaction. Therefore, Scavenger can retrieve keys corresponding to the hot-write data from DropCache over a recent period by recording keys merged during compaction and utilizing the LRU algorithm. Despite this, due to the small fraction of hot data and its sole role in hotspot identification, DropCache requires minimal memory resources and does not serve any foreground requests. It only records keys (32B per KV) of hot data. For larger datasets, space-efficient probabilistic data structures like CuckooFilter \cite{fan2014cuckoo} could be considered to reduce memory overhead.

\textbf{Hotness-aware Flush and GC}.
Building upon DropCache, Scavenger implements a hotness-aware write strategy during the generation of value SSTables for Flush and GC operations. As depicted in Figure~\ref{fig:design1-all}, Scavenger verifies the data keys to be written via DropCache. If a key hits DropCache, the associated data, deemed as hot data, is written to the hot vSST. Otherwise, it is written to the cold vSST. Ultimately, Scavenger partitions the data into hot and cold vSSTs during flush and GC.

After partitioning, the hot vSSTs, which contain more hot-write data, show a higher garbage ratio than the cold vSSTs over time. Our current GC strategy, triggered by the garbage ratio threshold, prioritizes files with higher garbage ratios for GC. This increases the chances of hot vSSTs for GC, improving GC efficiency and reducing unnecessary GC on cold data. Notably, the size of DropCache does not significantly affect the efficiency of GC, as we only need a sufficient difference in the garbage ratio between hot and cold vSSTs rather than a large amount of hot data or precise hotness.

\subsection{Space-aware Compaction based on Compensated Size}

\begin{figure}[tbp]
        \setlength{\abovecaptionskip}{5pt}
        \setlength{\belowcaptionskip}{-0.6cm}
	\centering
	\subfloat[Vanilla file size]{
            \includegraphics[width=.33\columnwidth]{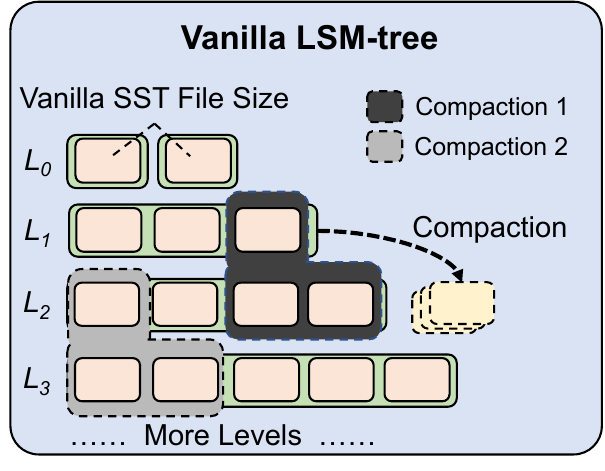}
            \label{fig:des2-comp-va-file-size}
        }\hspace{-3mm}
        \subfloat[kSST file size]{
            \includegraphics[width=.273\columnwidth]{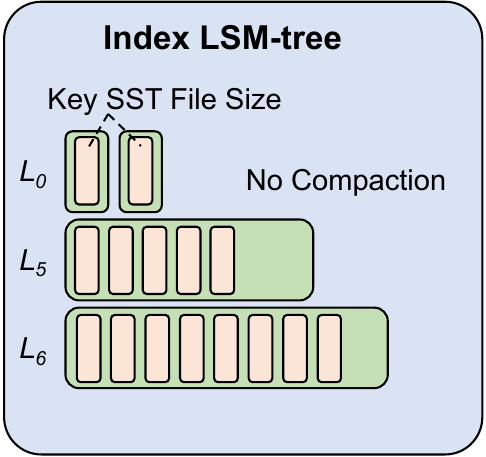}
            \label{fig:des2-comp-file-size}
        }\hspace{-3mm}
        \subfloat[Compensated file size]{
            \includegraphics[width=.33\columnwidth]{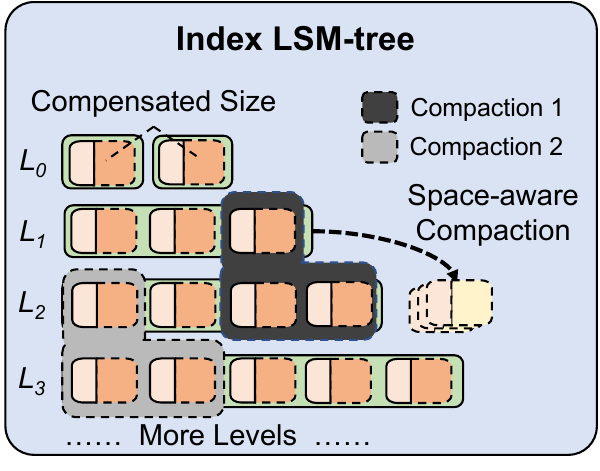}
            \label{fig:des2-comp-csize}
        }
	\caption{Different compaction strategies.}
\end{figure}

As discussed in Section \uppercase\expandafter{\romannumeral2}-D, the index LSM-tree contributes to space amplification in KV-separated LSM-trees as well. Figure~\subref*{fig:des2-comp-file-size} shows the main cause of severe space amplification in the index LSM-tree is the smaller file size written to it compared to the vanilla LSM-tree, which leads to delayed compaction. Delayed compaction presents two issues: it causes data accumulation at the upper level, which delays the prompt recycling of corresponding garbage, and it reduces the number of levels, obstructing parallel compaction execution. Space amplification results from the accumulation of upper-level data, and the slowed pace of space reclamation is due to the inability to execute compaction in parallel. Therefore, it is crucial for KV-separated LSM-trees to promptly schedule compaction and construct a multi-level LSM-tree structure to maintain the ideal space amplification of the index LSM-tree.

A straightforward approach would be to set smaller compaction trigger thresholds to facilitate more frequent compaction. However, determining suitable thresholds and adapting to workload fluctuations present significant challenges. Vanilla LSM-trees, such as RocksDB, shown in Figure~\subref*{fig:des2-comp-va-file-size}, utilize the leveled compaction strategy and dynamic level \cite{dynamic-level} to curtail space amplification. However, these strategies perform poorly in the KV-separated LSM-tree due to fluctuations in file size. Consequently, we propose a space-aware compaction strategy based on compensated size to achieve equivalent space amplification to the vanilla LSM-tree.

\textbf{Compensated Size-based compaction strategy} considers the original data size before scheduling the compaction. As depicted in Figure~\subref*{fig:des2-comp-csize}, we incorporate the value data related to the kSST file into its size calculation in the index LSM-tree, thereby converting a separated LSM-tree into a non-separated one via compensation. By recording the actual entry size, like BlobDB and Titan, or dependencies between kSST and vSST, similar to TerarkDB, in the kSST, we can promptly discern the corresponding native SST data size during compaction. The compensated size dynamically adapts based on the actual value size, accommodating workloads with diverse entry sizes, irrespective of static configurations. For the compaction score, we consider the sum of the compensated sizes of all files at each level as the compensated size for that level. A larger compensated size at a level increases the likelihood of initiating compaction to reclaim storage space in the index LSM-tree, thereby ensuring the maintenance of a strictly hierarchical structure within the index LSM-tree with fixed multipliers.

Besides using the compensated size for compaction level selection, it also guides the choice of compaction files at each level, aiming to push down high-density files swiftly. This leads to the quick conversion of hidden garbage linked to that file into exposed garbage, thus boosting the efficiency of GC. Compensation allows the index LSM-tree to swiftly mirror the multi-level structure of the vanilla LSM-tree and build more levels. More levels imply fewer data dependencies between compaction jobs, enhancing the parallelism of the compaction and speeding up space reclamation, as illustrated in Figure~\subref*{fig:des2-comp-csize}. With this multi-level structure and a fixed inter-level ratio (default 10), the space amplification of the index LSM-tree can converge to an approximate value of 1.11x, similar to vanilla LSM-trees, as shown in Figure~\subref*{fig:test6-features-index-sa}, thereby reducing the overall space amplification.

\subsection{Discussion}
\textbf{Applicability}. Though we implemented Scavenger based on TerarkDB, our solution is highly portable across other KV-separated LSM-trees. The new table structures, DTable and RTable, are modifications of RocksDB's BlockBasedTable. Hotness-aware Flush and GC do not rely on a specific implementation. For the compaction strategy based on compensated size, if the entry size is stored in the index LSM-tree as in BlobDB and Titan, we can directly calculate the compensated size and store an 8B total size in each SST metadata area for easy access, which introduces negligible overhead. TerarkDB already maintains the dependent vSST file numbers and corresponding entry counts in kSST. Therefore, we can directly calculate the compensated size based on the proportion of entries occupied by dependent data and the corresponding vSST file size. This can be fully implemented using the existing mechanism without extra space overhead.

\textbf{Availability}. Our tests in Section \uppercase\expandafter{\romannumeral2}-C revealed severe space amplification in existing KV-separated LSM-trees. However, storage space for a single instance of LSM-tree is limited in production environments. KVS cannot provide services once it exhausts storage space. We introduced a space-aware throttling mechanism to ensure service availability. As space nears full capacity, the strategy slows or halts foreground writes, lowering the garbage ratio threshold for aggressive GC. Foreground writing can resume after space reclamation, enhancing availability. Space-aware throttling effectively limits space use but can compromise performance, especially if KVS does not reclaim space timely. We set a maximum space threshold based on the available space quota to trigger throttling, providing a trade-off between space and performance.

\section{Evaluation}
In this section, we conduct evaluations of Scavenger, comparing it with several state-of-the-art KV-separated LSM-trees: BlobDB, Titan, and TerarkDB. We also employ the widely adopted LSM-tree, RocksDB (v8.2.1), as our baseline.

\subsection{Setup}
\textbf{Testbed}. Our test environment consists of a system with a 32-core Intel(R) Xeon(R) Silver 4314 CPU @ 2.40GHz and 64GB DDR4 memory. We employed a 500G KIOXIA NVMe SSD. The operating system kernel for the server is Linux 5.4, with an operating system version of Debian 10. We formatted the NVMe SSD using the ext4 file system.

\textbf{System Configuration}. Following the tuning guide \cite{rocksdb-tuning}, we standardized the configurations for all evaluated KVSs. The key-value separation threshold was set to 512B, the Memtable size to 64MB, the key Sorted String Tables (kSST) size to 64MB, the value Sorted String Tables (vSST) size to 256MB, and bloom filters to 10 bits/key. Both the foreground and background threads are configured to 16, with a 1GB BlockCache (about 1\% of the 100GB dataset). Our system enables direct I/O for background flush and compaction to mitigate the impact of page cache. Simultaneously, we disable readahead for GC operations of all KV-separated LSM-trees by default to ensure fairness \cite{blobdb, titan}. The garbage ratio threshold was set at 0.2 to regulate the frequency of GC.

\textbf{Fair Comparison}. It is unfair to evaluate different KV-separated LSM-trees without limiting space usage. We set a space threshold based on space-aware throttling to limit the maximum space available for a single KVS, ensuring fair performance comparison and limited costs. We set 1.5x the dataset size as the space limit unless otherwise specified.

\textbf{Workload}. We support fixed and variable-length workloads using a modified dbbench (from RocksDB) and YCSB-C \cite{ycsbc}, a C++ version of YCSB \cite{cooper2010benchmarking}. The key size was set to a constant 24B. For value lengths, we employ variable-length workloads, where a 1:1 ratio of small (uniformly distributed from 100B to 512B) to large (16KB) values, simulating the ByteDance OLTP database pattern, simplified as Mixed-8K (average size is about 8KB). Besides, we consider variable-length workloads with value sizes that conform to a generalized Pareto distribution \cite{hosking1987parameter, rocksdb-trace} and the average value size is about 1KB \cite{li2021differentiated}. For key distribution, we employ Zipfian distribution \cite{cooper2010benchmarking}, typical in real-world hotspot scenarios.

\begin{figure*}[tbp]
        \setlength{\abovecaptionskip}{5pt}
        \setlength{\belowcaptionskip}{-0.7cm}
	\centering
	\subfloat[Throughput under Mixed-8K]{
            \includegraphics[width=.29\textwidth]{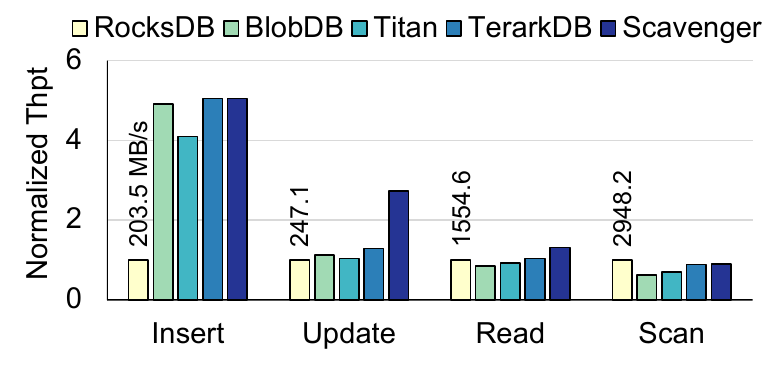}
            \label{fig:test1-micro-mixed-iops}
        }
        \subfloat[Throughput under Pareto-1K]{
            \includegraphics[width=.33\textwidth]{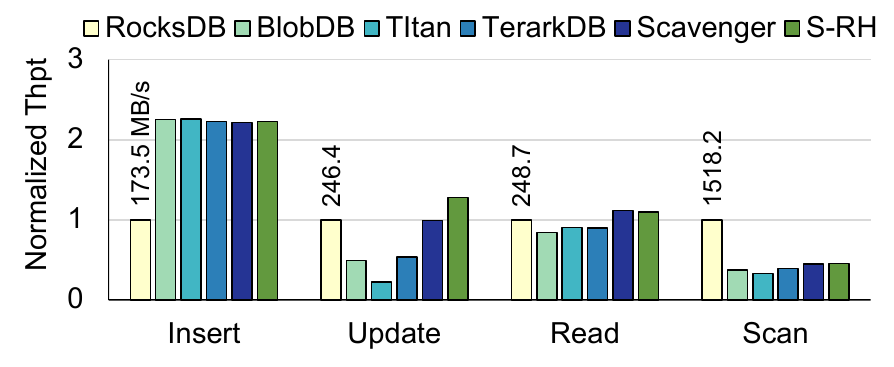}
            \label{fig:test1-micro-pareto-iops}
        }
        \subfloat[Disk I/O under Mixed-8K Update]{
            \includegraphics[width=.29\textwidth]{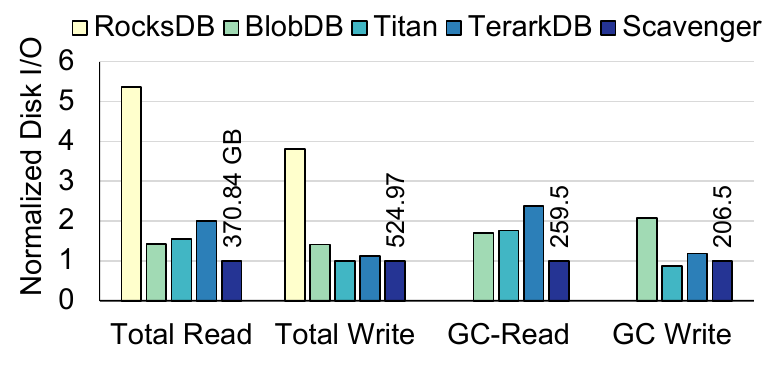}
            \label{fig:test1-micro-ndb-io}
        }
        \caption{Microbenchmarks under Mixed-8K and Pareto-1K with 1.5x space limit (\textit{\textbf{S-RH} denotes Scavenger-Readahead}). }
\end{figure*}

\begin{figure*}[tbp]
        \setlength{\abovecaptionskip}{5pt}
        \setlength{\belowcaptionskip}{-0.9cm}
	\centering
	\subfloat[Throughput under Mixed-8K]{
            \includegraphics[width=.40\textwidth]{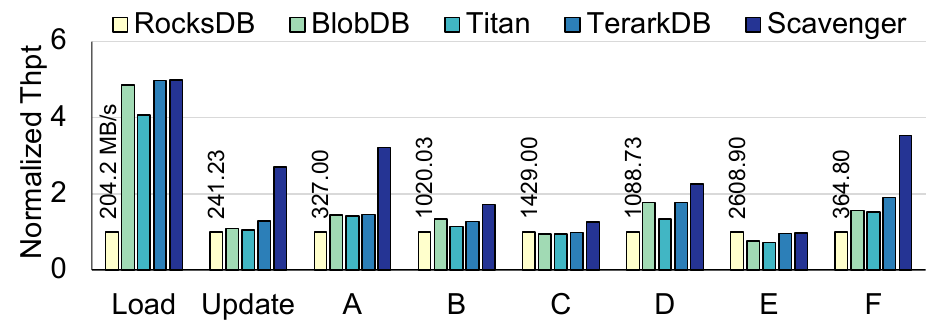}
            \label{fig:test3-ycsb-mixed}
        }
        \subfloat[Throughput under Pareto-1K]{
            \includegraphics[width=.40\textwidth]{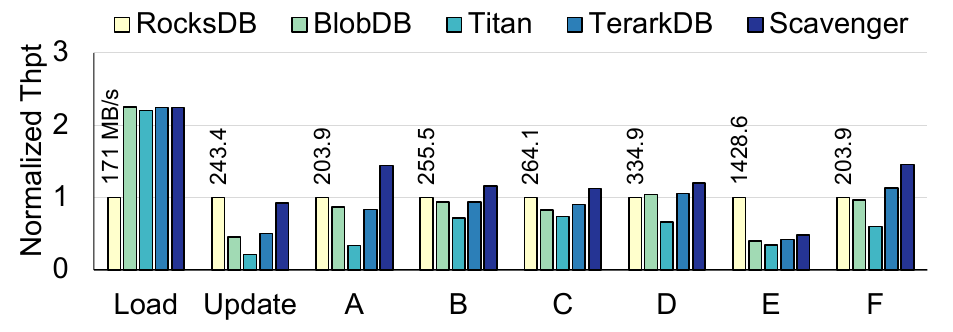}
            \label{fig:test3-ycsb-pareto}
        }
	\caption{YCSB under Mixed-8K and Pareto-1K with 1.5x space limit.}
\end{figure*}

\subsection{Microbenchmarks}
We evaluated the throughput of various key-value operations across different workloads (Mixed-8K and Pareto-1K), including inserting 100GB of key-value pairs, updating 300GB of pairs, reading 300GB, and performing 40 million range query requests. For scan workloads, scan lengths uniformly distributed from 1 to 1000. The testing procedure involved randomly loading 100GB of data and executing the update workload to generate substantial garbage data, thereby triggering GC operations. Subsequently, read and scan workloads were performed. To explore the trade-offs between space usage and time efficiency in existing schemes, we evaluated performance under a maximum space usage of 1.5 times the dataset size (150GB storage quota), and also examined performance and space amplification without any space constraints.

\textbf{Mixed-8K workload}. Figure \subref*{fig:test1-micro-mixed-iops} shows Scavenger significantly improves update performance, surpassing RocksDB by 2.7x, and BlobDB, Titan, TerarkDB by 2.4x, 2.6x, and 2.1x, respectively. It also exhibits 1.3x read performance over RocksDB due to prioritized caching of index data speeding up large-value access. Besides, Scavenger's insertion and scan performance are comparable to TerarkDB's. For insertion, all KV-separated LSM-trees perform 4-5x better than RocksDB. For scan, frequent compactions in RocksDB enhance value order, outperforming KV-separated LSM-trees. Notably, Scavenger and TerarkDB outperform others, as TerarkDB's efficient GC operations enhance data order, improving scan performance. As only update workload affects space amplification, Scavenger attains the most significant performance gain in update operations with limited space while preserving TerarkDB's performance advantages in other operations.

Considering the significant enhancement in Scavenger's update performance, we evaluated its read and write I/O impact under the update workload, as depicted in Figure \subref*{fig:test1-micro-ndb-io}. Compared to other KV-separated LSM-trees, Scavenger reduces read I/O by 42\% to 99\%. In terms of write I/O, Scavenger achieves a 12\% to 41\% reduction. Furthermore, we also examined the I/O associated with GC operations within KV-separated LSM-trees. This analysis reveals that the reduction in Scavenger's disk read and write I/O is primarily attributed to the minimization of I/O during GC operations.

\textbf{Pareto-1K workload}. Figure \subref*{fig:test1-micro-pareto-iops} shows that Scavenger significantly improves update performance, outperforming other KV-separated LSM-trees, BlobDB, Titan, and TerarkDB by 2.0x, 4.5x, and 1.9x, respectively. Compared to RocksDB, Scavenger's update performance is similar. However, unlike RocksDB, which enables readahead for compaction, all KV-separated LSM-trees disable readahead for GC. Therefore, we enabled readahead in Scavenger, denoted as \textbf{S-RH}, for further tests. We found that \textbf{S-RH} outperformed RocksDB in update performance by 1.28x. Additionally, Scavenger's insertion and read performances align with the conclusions drawn from the Mixed-8K workload, where Scavenger also performed the best. Regarding scan, Scavenger still outperforms other KV-separated LSM-trees but falls short of RocksDB, which sacrifices insert performance to improve data ordering.
\begin{figure}[tbp]
        \setlength{\abovecaptionskip}{5pt}
        \setlength{\belowcaptionskip}{-0.6cm}
	\centering
	\subfloat[Write Throughput]{
            \includegraphics[width=.44\columnwidth]{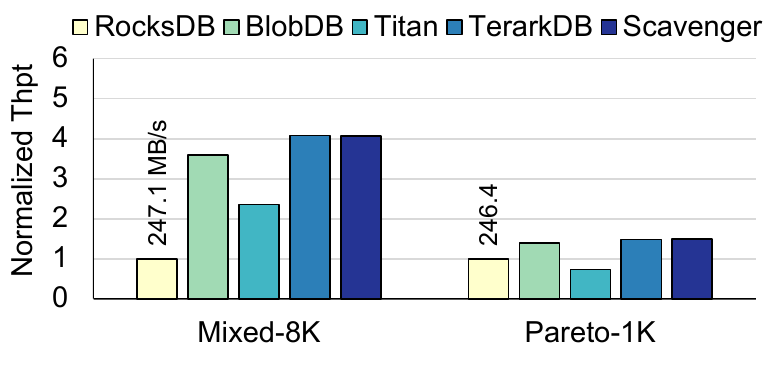}
            \label{fig:test2-micro-nolimit-update}
        }
        \subfloat[Space amplification]{
            \includegraphics[width=.44\columnwidth]{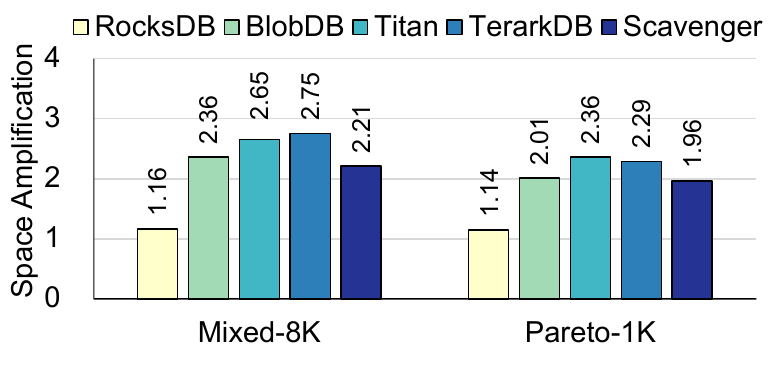}
            \label{fig:test2-micro-nolimit-space}
        }
        \label{fig:test2-micro-nolimit}
	\caption{Microbenchmarks without space limit.}
        \label{test2-micro}
\end{figure}

\textbf{Without space limit}. We assessed the performance and space amplification under Mixed-8K and Pareto-1K without any space limitations. Since only write operations influence space size, we mainly focused on the update workload's performance and space. As depicted in Figure \ref{test2-micro}, Scavenger delivers the best performance under both workloads, mirroring the foreground performance advantages of TerarkDB. Simultaneously, Scavenger shows a space amplification of 2.21 and 1.96, which is up to 40\% lower than that of other KV-separated LSM-trees, achieving a superior space-time trade-off.

\subsection{YCSB Evaluation}
We evaluated the performance of Scavenger using YCSB workloads \cite{cooper2010benchmarking}, with a focus on throughput after numerous updates. For each YCSB workload run, we adhered to the following steps: initializing 100GB of uniformly random data and then applying 300GB of updates to ensure that GC operations would run in all KV-separated LSM-trees. Subsequently, we executed YCSB workload A-F on the updated 300GB dataset, maintaining the same configuration as previously described.

\textbf{Mixed-8K workload}. Initially, we evaluated performance under Mixed-8K workloads, shown in Figure \ref{fig:test3-ycsb-mixed}. The results showed improved performance of write-intensive workloads with Scavenger. Specifically, compared with other KV Stores, Scavenger showed an improvement of approximately 2.2x-3.2x in the YCSB-A mixed-read-write (1:1) workload and 1.9x-3.5x in the YCSB-F workload. In scenarios with read-intensive workloads, Scavenger outperforms its competitors. Under the YCSB-E workload, Scavenger and TerarkDB demonstrate performance comparable to RocksDB, surpassing other KV-separated LSM-tree solutions.

\textbf{Pareto-1K workload}. We conducted evaluations under the Pareto-1K workload. Figure \ref{fig:test3-ycsb-pareto} shows that Scavenger performs best in all workloads except for YCSB-E. Scavenger performs better in write-intensive workloads, such as YCSB-A and YCSB-F, as with the previous test results. For the YCSB-E, RocksDB sacrifices write performance to improve data order; this is particularly effective since the key-value pairs are smaller under the Pareto distribution and more affected by data ordering, allowing it to outperform all KV-separated LSM-trees. Scavenger outperforms other KV-separated LSM-trees due to its increased GC operations.

\begin{figure}[t]
        \setlength{\abovecaptionskip}{5pt}
        \setlength{\belowcaptionskip}{-0.6cm}
	\centering
	\subfloat[Throughput]{
            \includegraphics[width=.44\columnwidth]{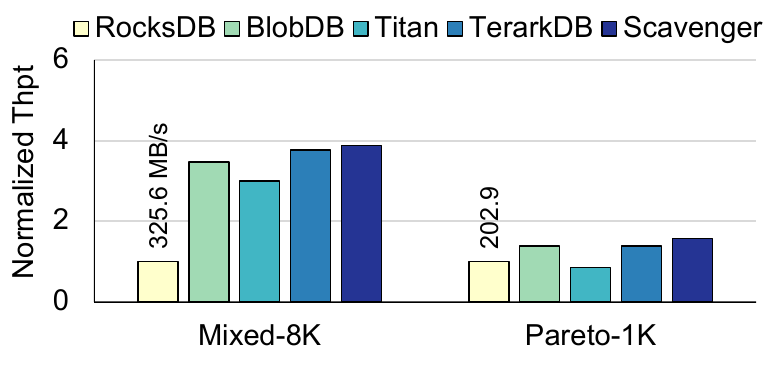}
            \label{fig:test4-ycsb-nolimit-a}
        }
        \subfloat[Space amplification]{
            \includegraphics[width=.44\columnwidth]{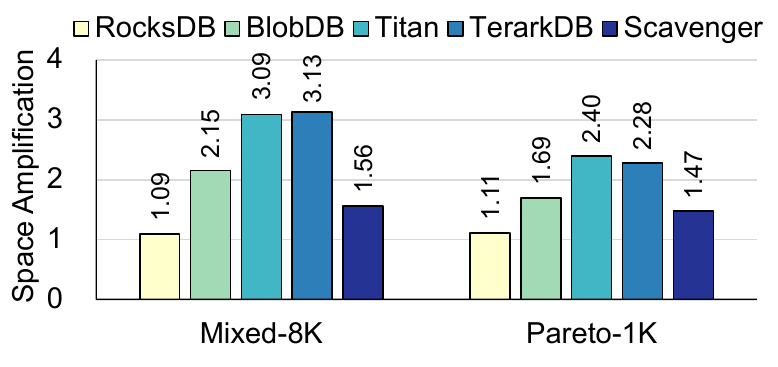}
            \label{fig:test4-ycsb-nolimit-space}
        }
	\caption{YCSB-A without space limit.}
        \label{test4-ycsb-nolimit}
\end{figure}

\textbf{Without space limit}. In the previous section, we evaluated the performance and space amplification of update workloads without any space limitations. In this section, we focus on write-intensive workloads, particularly those closely related to space amplification, as represented by YCSB-A. Figure \ref{test4-ycsb-nolimit} presents the results for Mixed-8K and Pareto-1K workloads. The conclusions drawn here are similar to those from the microbenchmark tests, with Scavenger showing the best foreground performance and significantly lower space amplification than other KV-separated LSM-trees.

\subsection{Scavenger Features}
\begin{figure}[tbp]
        \setlength{\abovecaptionskip}{5pt}
        \setlength{\belowcaptionskip}{-0.8cm}
	\centering
	\subfloat[Compaction and GC Features]{
            \includegraphics[width=.68\columnwidth]{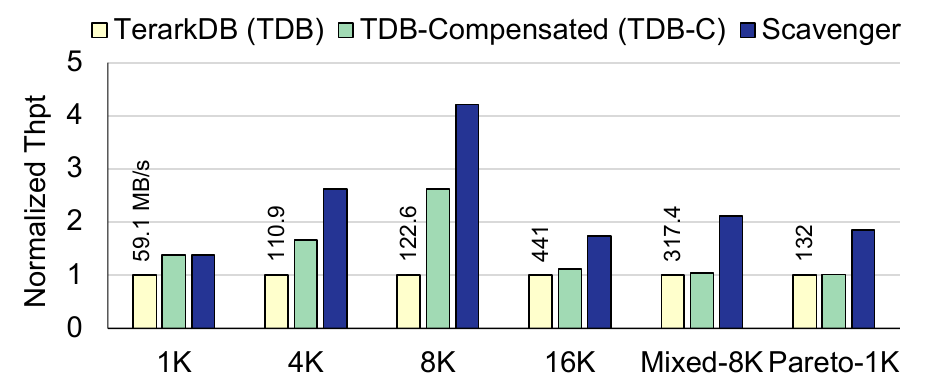}
            \label{fig:test5-features-all}
        }
        \subfloat[GC Features (RWL)]{
            \includegraphics[width=.30\columnwidth]{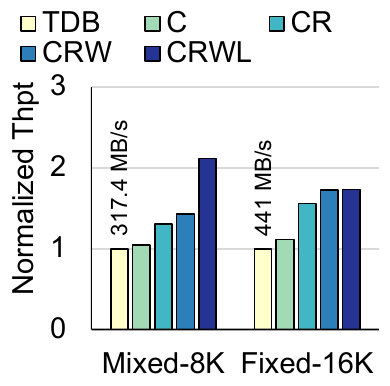}
            \label{fig:test5-features-gc}
        }
	\caption{Write throughput of features under 1.5x space limit.}
         \label{fig:test5-features}
\end{figure}

\begin{figure}[tbp]
        \setlength{\abovecaptionskip}{5pt}
        \setlength{\belowcaptionskip}{-0.6cm}
	\centering
	\subfloat[Compaction and GC Features]{
            \includegraphics[width=.68\columnwidth]{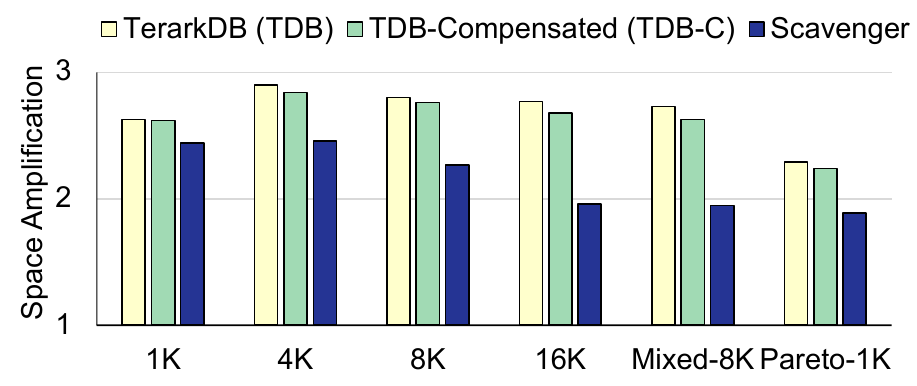}
            \label{fig:test6-features-all}
        }
        \subfloat[GC Features (RWL)]{
            \includegraphics[width=.30\columnwidth]{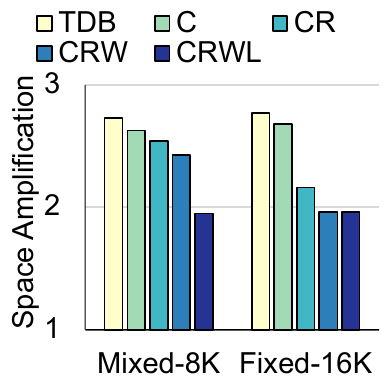}
            \label{fig:test6-features-gc}
        }
	\caption{Space amplification of features without space limit.}
         \label{fig:test6-features-1}
\end{figure}

\textbf{Limited space}. We evaluated the impact of different optimization techniques by conducting tests on the space-aware compaction strategy and I/O-efficient GC in TerarkDB (\textbf{TDB}). \textbf{TDB-C} integrates the space-aware compaction strategy in TDB, while \textbf{Scavenger} utilizes I/O-efficient GC in TDB-C. Figure \subref*{fig:test5-features-all} demonstrates the space-aware compaction strategy effectively accelerating space recycling and improving update performance by 1.6x-2.6x under fixed-length workloads. The performance benefits of the space-aware compaction strategy are less apparent for variable-length workloads, as the index LSM-tree stores numerous key-value pairs, leading to a significant increase in file size and a multi-level structure. Concurrently, I/O-efficient GC proves effective in most scenarios, especially for large-value and variable-length workloads. Figure \subref*{fig:test5-features-gc} illustrates that lazy read (\textbf{R}) significantly improves performance under fixed-length workloads, while GC-Lookup (\textbf{L}) optimization excels under variable-length workloads.

\begin{figure}[tbp]
        \setlength{\abovecaptionskip}{5pt}
        \setlength{\belowcaptionskip}{-0.6cm}
	\centering
	\subfloat[Space amplification of index]{
            \includegraphics[width=.49\columnwidth]{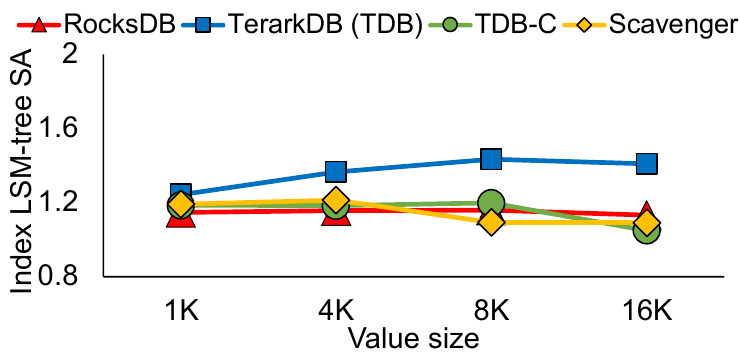}
            \label{fig:test6-features-index-sa}
        }
        \subfloat[Exposed garbage of value]{
            \includegraphics[width=.49\columnwidth]{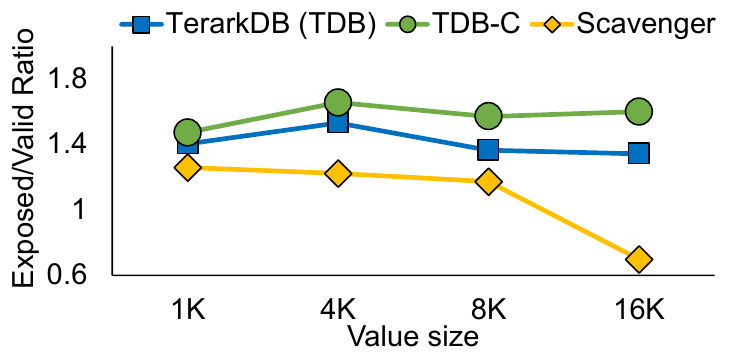}
            \label{fig:test6-features-exposed-sa}
        }
	\caption{Sources of space amplification without space limit.}
        \label{test6-features}
\end{figure}

\textbf{Without space limit}. We evaluated write performance and space amplification without space constraints. Scavenger, TerarkDB, and TDB-C showed similar write performance. Our focus was on space amplification. As shown in Figure~\subref*{fig:test6-features-all}, enabling space-aware compaction alone reduces space amplification by up to 4\%. Given these results, enabling I/O efficient GC is crucial to achieve a reduction in space amplification by up to 30\%. Figure~\ref{test6-features} provides a more in-depth root cause analysis. Space-aware compaction reduces index LSM-tree's space amplification to nearly 1.1 but increases exposed garbage in value data. Without GC optimization, the system cannot promptly recycle the substantial amount of generated garbage, leading to a negligible reduction in space amplification. Furthermore, we tested the effects of different GC optimization techniques, as shown in Figure~\subref*{fig:test6-features-gc}. GC-Lookup optimization (\textbf{L}) has the most significant effect on variable-length workloads, while for fixed-length workloads, the effects of lazy read optimization (\textbf{R}) are the most significant. Write optimization (\textbf{W}) benefits all workload types in our tests.

\subsection{Performance under Different Workloads and Limits}
\subsubsection{Varying value size}
\begin{figure*}[tbp]
        \setlength{\abovecaptionskip}{5pt}
        \setlength{\belowcaptionskip}{-0.6cm}
	\centering
	\subfloat[Throughput under Fixed Workloads]{
            \includegraphics[width=.34\textwidth]{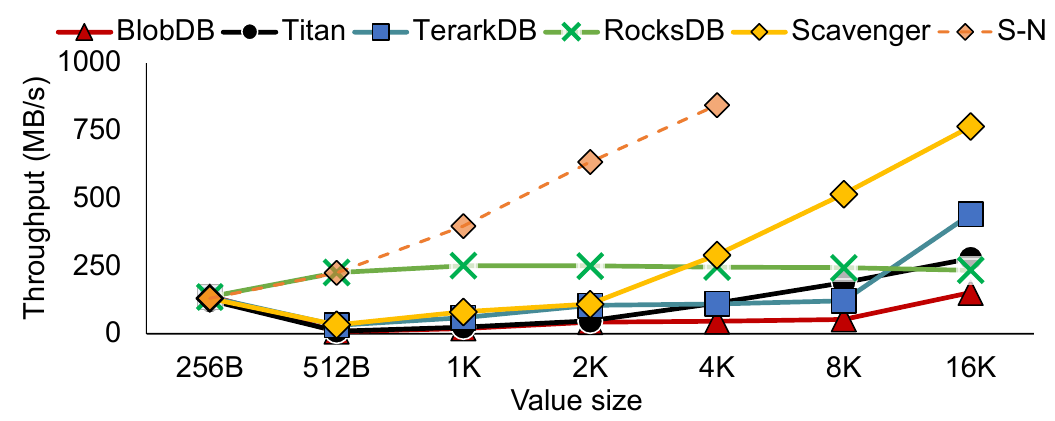}
            \label{fig:test7-diff-fixed-size}
        }
	\subfloat[Throughput under Mixed Workloads]{
            \includegraphics[width=.3\textwidth]{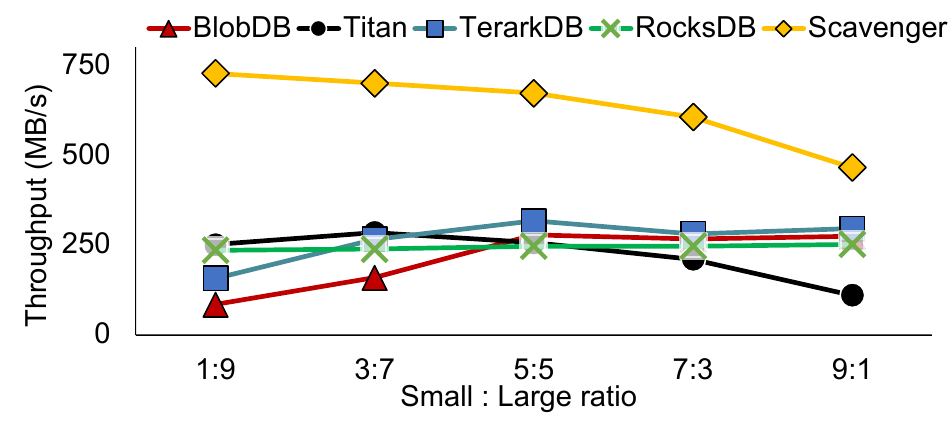}
            \label{fig:test7-diff-mixed-size}
        }
        \subfloat[Throughput under varying skewness]{
            \includegraphics[width=.3\textwidth]{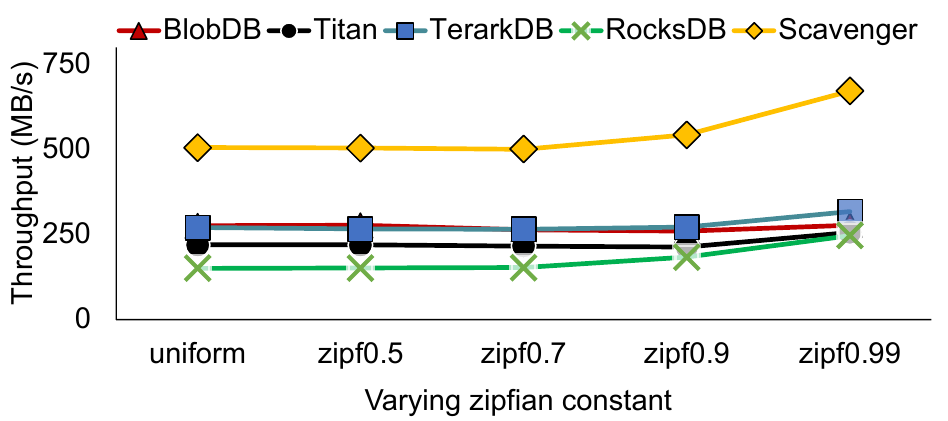}
            \label{fig:test8-diff-skew}
        }
	\caption{Update performance under varying workloads with 1.5x space limit (\textit{\textbf{S-N} denotes Scavenger without space limits}).}
\end{figure*}

To assess the adaptability of Scavenger across diverse workloads, we evaluated its write performance under varying value-size distribution workloads.


\textbf{Fixed size}. As shown in Figure \subref*{fig:test7-diff-fixed-size}, we evaluated various value sizes ranging from 256B to 16KB. We observed a decline in the performance of all KV-separated LSM-trees between 256B and 2KB, falling below RocksDB's. This implies that KV-separated LSM-trees struggle to balance high performance and reasonable space amplification for small-value workloads. From prior analysis, the readahead mechanism significantly affects performance, particularly for small-value workloads ($\leq 2KB$). Increasing the separation threshold or enabling readahead can mitigate this issue. However, Scavenger still outperformed other KV-separated LSM-trees by 1.1x-4.0x. Concurrently, we evaluated Scavenger under 512B-4KB without space throttling (\textbf{S-N}) and found that it surpassed RocksDB, despite a 1.3x increase in space usage.

\textbf{Mixed size}. We evaluated the Mixed workload by altering the proportions of small and large key-value pairs, with 1:9, 3:7, 5:5, 7:3, and 9:1, as depicted in Figure \subref*{fig:test7-diff-mixed-size}. Small values followed a uniform distribution with sizes varying from 100 to 512B, while large values remained consistently at 16KB. The analysis revealed that as the proportion of small values escalated, Scavenger's performance advantage diminished due to the index LSM-tree storing more data. Despite the decrease, Scavenger still surpassed other KVS solutions. Its performance continued to be superior, particularly in scenarios where large values constituted a substantial portion.

\subsubsection{Varying skewness}
We evaluated Scavenger's update performance under different workload skewness with varying Zipfian constants. A higher constant indicates a more skewed workload. Uniform workloads were also tested. Figure \subref*{fig:test8-diff-skew} shows that Scavenger consistently outperforms other key-value stores across all workloads. As the skewness of the workload escalates, Scavenger's performance improves significantly. With a Zipfian constant of 0.99, Scavenger exhibits a significant improvement of 2.1x-2.7x. This is attributable to Scavenger's ability to exploit the advantages of data distribution with concentrated hotspots effectively.

\subsubsection{Varying space limit}
\begin{figure}[tbp]
    \setlength{\abovecaptionskip}{5pt}
    \setlength{\belowcaptionskip}{-0.55cm}
    \centering
    \includegraphics[width=0.4\textwidth]{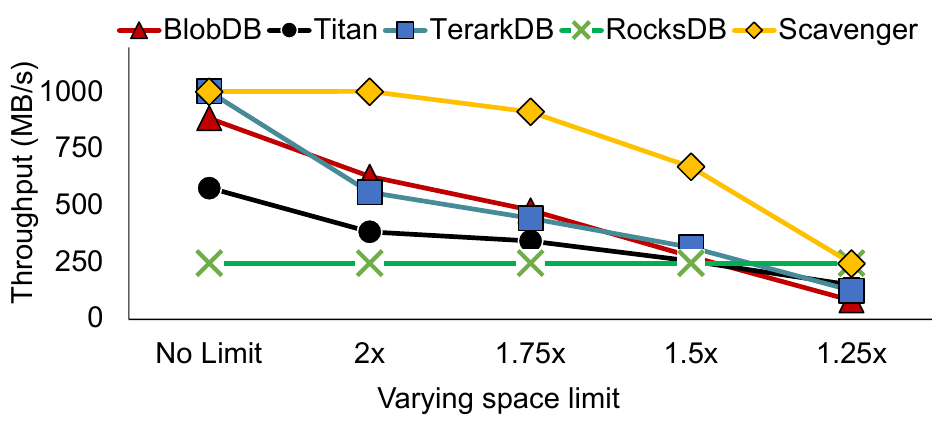}
    \caption{Update performance under varying space limits.}
    \label{fig:test9-diff-space}
\end{figure}

Figure~\ref{fig:test9-diff-space} shows Scavenger's performance under different space constraints. We set space-aware throttling thresholds at 1.25x, 1.5x, 1.75x, and 2x dataset sizes. The no-limit scenario with disabled throttling was also considered. As space constraints ease and quotas increase, the performance advantages of KV-separated LSM-trees become clearer. RocksDB's performance, which benefits from low space amplification, remains relatively stable. Scavenger consistently shows superior performance under more stringent quotas, such as 1.25x and 1.5x. Within the 1.25x space limit, only Scavenger achieves throughput equivalent to RocksDB across all KV-separated LSM-trees. It implies that the benefits of KV separation diminish under strict space constraints. However, Scavenger's ability to minimize GC-related I/O, enhance space reclamation, and synchronize it with foreground writes results in performance comparable to RocksDB's. We also tested under the Pareto-1K workload, yielding similar conclusions, so results are omitted for brevity.

\subsection{Space overhead}
\begin{table}[htp]
    \setlength{\abovecaptionskip}{5pt}
    \setlength{\belowcaptionskip}{-0.4cm}
    \centering
    \caption{Space usage under varying workloads for Insert}
    \begin{tabular}{|c|c|c|c|c|c|} \hline  
         \textbf{Space (GB)}&  \textbf{1K}&  \textbf{4K}&  \textbf{16K}&  \textbf{Mixed-8K}& \textbf{Pareto-1K}
\\ \hline  
         TerarkDB&  107.30&  102.71&  100.77&  100.83& 101.91
\\ \hline  
         Scavenger&  112.43&  103.23&  100.81&  101.12& 104.14
\\ \hline  
         Ratio&  4.78\%&  0.51\%&  0.04\%&  0.29\%& 2.19\%\\ \hline 
    \end{tabular}
    
    \label{tab:overhead}
\end{table}
Scavenger incurs minor space overhead due to additional indexes. To mitigate the impact of GC on space, we evaluated the space overhead under an insert workload. Table~\ref{tab:overhead} presents the space usage of Scavenger with varying value sizes. Scavenger incurs less than 5\% additional space compared to TerarkDB, a nearly negligible amount, particularly for large-value workloads. The additional space overhead stems from the indexes RTable introduces for each key-value pair, while both DTable and the compensated-size-based compaction fully utilize the existing storage layout, incurring no extra overhead.

\section{Related Work}
Numerous studies have been conducted to enhance the read and write performance of LSM-tree-based KVS.

\textbf{Compaction Optimization}. Efforts to enhance LSM-tree performance target compaction overhead by relaxing orderliness restrictions. PebblesDB \cite{raju2017pebblesdb} relaxes single-level global orderliness, adopting FLSM structure to reduce write amplification, despite increasing read and space amplification. Other schemes, such as UniKV \cite{zhang2020unikv}, WipDB \cite{zhao2021wipdb}, and REMIX \cite{zhong2021remix}, utilize partitioning to minimize compaction overhead, while BlockDB \cite{wang2022reducing} and B+LSM \cite{liu2023closing} optimize compaction granularity and strategy to reduce write amplification. L2SM \cite{huang2021less} and LSbM \cite{teng2017lsbm} use buffers to achieve an effect similar to tiering compaction. However, they also experience the drawbacks of read and space amplification associated with tiering compaction. Sarkar et al. \cite{sarkar2021constructing, sarkar2022compactionary} introduced a design space for compaction in vanilla LSM-trees, but they neglected the issue of space amplification in KV-separated LSM-trees. Although some compaction strategies may help reduce the space amplification of vanilla LSM-trees, they cannot be effectively applied to KV-separated LSM-trees because they cannot perceive the size of the separated values.

\textbf{Key-Value Separation}. KV separation has gained industry recognition for reducing LSM-tree read and write amplification. WiscKey \cite{lu2017wisckey} and HashKV \cite{chan2018hashkv} first explored KV-separation but overlooked query performance, particularly for range queries. Despite implementing learning indexes atop WiscKey by Bourbon \cite{dai2020wisckey} to improve read performance, it did not fundamentally alter the unordered storage layout. TerarkDB \cite{terarkdb}, Titan \cite{titan}, DiffKV \cite{li2021differentiated}, and BlobDB \cite{rocksdb} have improved value orderliness by utilizing structures similar to the original SST. Despite this, they still experience substantial fluctuations in performance due to substantial GC overhead. Titan, in particular, is encountering increased contentions with foreground write operations due to the requirement for write-back during GC. The compaction-triggered GC strategy of DiffKV and BlobDB further couples compaction and GC, significantly diminishing GC space reclamation efficiency. In comparison, Scavenger significantly reduces GC overhead and accelerates space reclamation while incurring lower I/O costs. Consequently, Scavenger achieves a more favorable balance between performance and space amplification.

\section{Conclusion}
This paper proposes Scavenger, a KV-separated LSM-tree-based KVS that provides better space-space trade-offs. We analyze the root causes of space amplification in KV-separated LSM-trees and devise an I/O-efficient garbage collection scheme to minimize I/O overhead. We also integrate a space-aware, compensated-size-based compaction strategy to reduce the space amplification of index LSM-trees. Scavenger rapidly reclaims space without sacrificing foreground performance.

\section*{Acknowledgments}
This work, supported by NSFC (No. U22A2027 and 61832020) and Shenzhen Technology Scheme Project (JCYJ20210324141601005), acknowledges the insightful comments of anonymous reviewers and the technical assistance of Ming Zhao from ByteDance.



\clearpage
\newpage
\bibliographystyle{IEEEtran}
\bibliography{IEEEabrv, icde}

\end{document}